\documentclass[%
 aip,
 jmp,%
 amsmath,amssymb,
 reprint,%
]{revtex4-1}

\usepackage{graphicx}
\graphicspath{{figures_pdf/}}
\usepackage{dcolumn}
\usepackage{bm}
\usepackage{subfigure}

\usepackage{soul}
\usepackage{amsmath,amssymb,amsfonts}
\usepackage{graphics}
\usepackage{color}
\usepackage{xcolor}
\definecolor{rev1}{rgb}{0,0,0}

\usepackage{algorithm}
\usepackage{algorithmic}
\usepackage{enumitem}
\setlist[itemize]{leftmargin=*}
\setlist[enumerate]{leftmargin=*}

\usepackage{comment}
\usepackage{hyperref}
\usepackage{lipsum}
\usepackage{tabularx}
\usepackage{mathrsfs}
\usepackage{stackengine}
\newcommand{\overbar}[1]{\mkern 1.5mu\overline{\mkern-1.5mu#1\mkern-1.5mu}\mkern 1.5mu}

\begin{document}



\title{Data assimilation empowered neural network parameterizations for subgrid processes in geophysical flows}

\author{Suraj Pawar}
\affiliation{ 
School of Mechanical \& Aerospace Engineering, Oklahoma State University, Stillwater, OK 74078, USA.
}%

\author{Omer San}%
 \email{osan@okstate.edu}
\affiliation{ 
School of Mechanical \& Aerospace Engineering, Oklahoma State University, Stillwater, OK 74078, USA.
}%




\date{\today}

\begin{abstract}

In the past couple of years, there is a proliferation in the use of machine learning approaches to represent subgrid scale processes in geophysical flows with an aim to improve the forecasting capability and to accelerate numerical simulations of these flows. Despite its success for different types of flow, the online deployment of a data-driven closure model can cause instabilities and biases in modeling the overall effect of subgrid scale processes, which in turn leads to inaccurate prediction. To tackle this issue, we exploit the data assimilation technique to correct the physics-based model coupled with the neural network as a surrogate for unresolved flow dynamics in multiscale systems. In particular, we use a set of neural network architectures to learn the correlation between resolved flow variables and the parameterizations of unresolved flow dynamics and formulate a data assimilation approach to correct the hybrid model during their online deployment. We illustrate our framework in a set of applications of the multiscale Lorenz 96 system for which the parameterization model for unresolved scales is exactly known, \textcolor{rev1}{and the two-dimensional Kraichnan turbulence system for which the parameterization model for unresolved scales is not known a priori.} Our analysis, therefore, comprises a predictive dynamical core empowered by (i) a data-driven closure model for subgrid scale processes, (ii) a data assimilation approach for forecast error correction, and (iii) both data-driven closure and data assimilation procedures. We show significant improvement in the long-term prediction of the underlying chaotic dynamics with our framework compared to using only neural network parameterizations for future prediction. Moreover, we demonstrate that these data-driven parameterization models can handle the non-Gaussian statistics of subgrid scale processes, and effectively improve the accuracy of outer data assimilation workflow loops in a modular non-intrusive way.       


\end{abstract}


\maketitle


\section{Introduction}
Geophysical flows are characterized by the multiscale nature of flows where there is a massive difference between the largest and smallest scales, and these scales interact with each other to exchange heat, momentum, and moisture. This makes the numerical simulations of geophysical flows in which every flow feature is resolved computationally unmanageable, even though the physical laws governing these processes are well known. Therefore, the atmosphere and ocean models compute the approximate numerical solution on the computational grid that consists of $O(10^7)$ to $O(10^8)$ grids with a spacing of $O(10~\text{km})$ to $O(100~\text{km})$. The effect of unresolved scales is taken into account by using several parameterization schemes, which represent the dynamics of subgrid scale processes as a function of resolved dynamics \cite{stensrud2009parameterization,duan2007stochastic,randall1989cloud}. However, the weather projection is marred by large uncertainties in the parameters of these parameterization schemes, and also due to incorrect structure of these parameterizations equations itself   \cite{schneider2017earth, draper1995assessment,holloway2009moisture}.

Typically, the parameters of these parameterization schemes are estimated by the model tuning process based on the observations from experimental and field measurements or the data generated from high-resolution numerical simulations \cite{jakob2003improved,jakob2010accelerating}. The nonlinear and multiscale nature of geophysical flows makes this tuning procedure cumbersome and can impede accurate climate prediction \cite{zhao2016uncertainty}. A recent development in machine learning, particularly deep learning \cite{lecun2015deep}, along with the huge volume of data gathered from high-resolution numerical simulations \cite{neumann2019assessing} and remote sensors measurements \cite{kuenzer2014earth} offers an alternative to the physics-based parameterization schemes and can pave the way for improved climate and weather models. Deep learning approaches have been demonstrated to be successful for different scientific tasks in Earth system science, such as extreme weather pattern detection \cite{liu2016application}, precipitation nowcasting \cite{shi2017deep}, transport process modeling \cite{de2019deep}, and many more. Deep learning has also been utilized to represent subgrid scale processes in climate models. \citet{rasp2018deep} trained a deep neural network (DNN) to emulate a cloud resolving model and formulated a procedure to produce stable results in the online deployments close to the original super-parameterized global circulation model. \citet{gentine2018could} used an ensemble of random forests as a machine learning (ML) algorithm to parameterize the moist convection and implemented it in a global circulation model. They demonstrated the stable and robust performance of ML based parameterization in capturing important climate statistics including precipitation extremes. 

Along with the Earth system science, there is a surge in the application of machine learning for fluid mechanics. Readers are directed to an excellent review by \citet{brunton2019machine} on how ML algorithms are being used for augmenting the domain knowledge, automating tasks such as flow-control and optimization by the fluid mechanics' community. In a recent perspective, \citet{brenner2019perspective} discuss the strength and limitations of ML based algorithms to advance fluid mechanics. The closure problem in turbulence modeling is similar to the parameterization in climate modeling and is encountered in Reynolds-Averaged Navier-Stokes (RANS) and large eddy simulation (LES) which are widely adopted for engineering flow simulations. There have been several studies that use ML algorithms to address the turbulence closure problem \cite{duraisamy2019turbulence,sarghini2003neural,gamahara2017searching,maulik2019accelerating}. \citet{ling2016reynolds} proposed a novel neural network architecture with embedded Galilean invariance for the prediction of Reynolds stress anisotropy tensor. \citet{wang2017physics} employed random forest as an ML algorithm to reconstruct the discrepancy \textcolor{rev1}{in} RANS-modeled Reynolds stresses and evaluated its performance for fully developed turbulent flows and separated flows. Deep learning has also been utilized for LES of turbulent flows, for example, subgrid scale closure modeling of Kraichnan turbulence \cite{maulik2019subgrid}, decaying homogeneous isotropic turbulence \cite{beck2019deep}, forced isotropic turbulence \cite{xie2020modeling}, compressible isotropic turbulence \cite{xie2019artificial}, and wall-bounded turbulence \cite{srinivasan2019predictions}. The feasibility of deep learning has been investigated to produce a predictive model for turbulent fluxes, such as heat fluxes \cite{kim2020prediction} and anomalous fluxes in drift-wave turbulence \cite{heinonen2020turbulence}.
In a recent work, \citet{novati2020automating} introduced a multi-agent reinforcement learning framework as an automated discovery tool for turbulence models and applied it to forced homogeneous isotropic turbulence. Besides turbulence closure modeling, deep learning has been proved to be very successful for challenging problems such as super-resolution of turbulent flows \cite{fukami2019super,jiang2020meshfreeflownet,subramaniam2020turbulence}, data-driven modeling of chaotic systems \cite{pathak2018model,vlachas2020backpropagation,wan2018data}, reduced order modeling of high-dimensional multiphysics systems \cite{lee2019model,mohan2019compressed,qian2020lift,rahman2019non}, and developing forecast models for complex physical systems \cite{wang2019towards,szunyogh2020machine,cheng2020data}.

Despite the development of deep learning algorithms as a powerful tool to extract spatio-temporal patterns from the data, these methods are criticized for their black-box nature and are prone to produce physically inconsistent results due to their lack of generalizability \cite{faghmous2014theory,wagner2016theory}. Moreover, the increase in spatial and temporal dimensionalities raises a computational challenge in terms of the training. Hence, it is essential to integrate machine learning with physics-based modeling to address the challenge of interpretability, physical consistency, and computational burden \cite{reichstein2019deep}. One way to combine machine learning with physics-based modeling is by incorporating physical conservation laws into training through a regularization term added to the loss function of a neural network \cite{raissi2019physics,subramaniam2020turbulence,raissi2018hidden,wu2020enforcing,erichson2020shallow}. Another way is to change the structure of neural network architecture to enforce physical conservation laws as hard constraints \cite{mohan2020embedding,marquez2017imposing}. The hybrid modeling in which a sub-model within the physics-based model is replaced by machine learning methods is another approach to address the limitation of pure data-driven methods \cite{reichstein2019deep,de2019deep,karpatne2017theory}. One of the issues with hybrid models is that the trained neural network often suffers from instability once they are deployed in the forward model. For example, a small change in the training dataset or the input and output vector of the neural network led to unpredictable blow-ups in the global circulation model that employs a neural network to emulate cloud resolving model \cite{rasp2018deep,rasp2020coupled}. Similarly, \citet{brenowitz2019spatially} found that the nonphysical correlations learned by neural networks were the cause of instabilities in their online deployment within the global circulation model \cite{brenowitz2018prognostic} and developed an approach to ensure stability. \citet{wu2019reynolds} highlighted the gap between \textit{a priori} and \textit{a posteriori} performance of data-driven Reynolds stress closure models as the RANS equations with such model can be ill-conditioned. Therefore, even though data-driven turbulence closure models predicted better closure terms, their online deployment does not lead to significant improvement in the mean velocity field prediction \cite{gamahara2017searching,wang2017physics}. \citet{wu2019reynolds} proposed a metric to evaluate the conditioning of RANS equations in the \textit{a priori} settings and showed that the implicit treatment of Reynolds stresses leads to reduced error in mean velocity prediction.

Data assimilation (DA) is a well-established discipline where observations are blended with the model to take uncertainties into account for improving the numerical prediction of the system \cite{lewis2006dynamic,simon2006optimal,evensen2009data,xiao2018parameterised,zerfas2019continuous,arcucci2019optimal} and can be applied to achieve accurate prediction in hybrid models that employ data-driven model as a sub-model for some processes (for example subgrid scale processes). DA tools are being extensively utilized in geoscience and numerical weather forecast centers to correct background predictions based on a combination of heterogeneous measurement data coming from ground observations and satellite remote-sensing. These techniques have been also investigated recently for integrating experimental data into large-eddy simulations of engineering flows \citep{labahn2020ensemble,mons2016reconstruction,da2018ensemble,colburn2011state}. In a DA workflow, we merge forward model predictions with observational data. However, it has been often remarked that no-model is correct but some of them are useful. In typical DA studies and twin experiments, therefore, the subgrid scale processes have been modeled as Gaussian noise due to a lack of structural information on their mechanisms. If we would know their dynamics either structurally or functionally, for sure it would be wise to include them in the model before a DA analysis is executed. However, the subgrid scale processes in turbulent flows often cannot be accurately modeled by Gaussian noise, and ML methodologies can be adopted to get a grip on subgrid scale processes. Hence, we put forth a neural network based statistical learning approach to improve model uncertainty and incorporate this information as a data-driven closure term to the forward model. We examine how the forecast error reduces due by including ML based closure term to the underlying forward model. Indeed, the integration of DA with ML methodologies holds immense potential in various fields of physical science \cite{li2020harmonizing,lguensat2017analog,tang2019deep,bocquet2020bayesian,brajard2020combining} and we demonstrate this through our study.





In this work, we propose a neural network closure framework in developing hybrid physics-ML models supplemented with DA for multiscale systems. In particular, we advocate the use of sequential DA techniques to \textcolor{rev1}{improve the state estimate of the system} by incorporating observations into a model equipped with neural network parameterization schemes for unresolved physics. To this end, we use real-time measurements to regularize ML empowered predictive tools through ensemble Kalman filter based approach. \textcolor{rev1}{Our first example} a two-level Lorenz 96 model \cite{lorenz1996predictability} for our numerical experiments since it generates a controllable test case for advancing turbulence parameterization theories, especially in the age of data-driven models. The Lorenz 96 is an idealized model of atmospheric circulation and is used widely to test research ideas \cite{law2016filter,karimi2010extensive,herrera2011role,van2019symmetries}. Even though the dynamics of both large and small scales are known exactly for a two-level Lorenz 96 model, it is very difficult to predict it because of the strong interplay between fast and slow subsystems. Therefore, we select this multiscale model for the assessments of data-driven closures for capturing the physics of subgrid scales. Since we use an ``explicit" evolution equation for the closure parameterizations, we can easily assess the data-driven models in \emph{a posteriori} simulations. This often comprises a challenging task in LES computations since the low-pass filtering operation is ``implicitly" applied to the governing equations. \textcolor{rev1}{We further extend our framework to Kraichnan turbulence \cite{kraichnan1967inertial}, where it is shown that the DA improves the state estimate of the hybrid physics-ML model and this leads to better prediction for statistical properties like kinetic energy spectra and vorticity structure functions in comparison with high-fidelity direct numerical simulation (DNS).} Our approach is multifaceted in at least two ways. We first show that the infusion of the DA approaches improves the forecasting quality of predictive models equipped with data-driven parameterizations. Second, we also demonstrate that the data-driven parameterizations help significantly to reduce forecast errors in DA workflows. Therefore, our modular framework can be considered as a way to incorporate real-time observations that are prevalent in today's weather forecast station into hybrid models constituted from a physics-based model as the dynamical core of the system, and a data-driven model to describe unresolved physics.  

The paper is structured as follows. In Section~\ref{sec:ml96}, we discuss the problem of parameterizations using a two-level Lorenz 96 model \textcolor{rev1}{and Kraichnan turbulence} as a prototypical examples. Section~\ref{sec:nnp} details two types of neural network utilized in this study for learning the mapping between resolved variables and parameterizations of unresolved scales. We explain the methodology of sequential data assimilation and ensemble Kalman filter based algorithms in Section~\ref{sec:da}.
In Section~\ref{sec:results}, we discuss the findings of our numerical experiments with a two-level Lorenz 96 model \textcolor{rev1}{and Kraichnan turbulence}. Finally, we conclude with the summary and direction for future work in Section~\ref{sec:conclusion}.



\section{Parameterizations in multiscale systems} \label{sec:ml96}

\subsection{Two-level Lorenz 96 model}
In this section, we describe the two-level variant of the Lorenz 96 model proposed by \citet{lorenz1996predictability}. This model has been extensively investigated  to study stochastic parameterization schemes\cite{palmer2001nonlinear,wilks2005effects,crommelin2008subgrid}, scale-adaptive parameterizations\cite{vissio2018proof}, and neural network parameterizations\cite{rasp2020coupled,wikner2020combining}. The two-level Lorenz 96 model can be written as 
\begin{align}
    \frac{d X_i}{dt} &= -X_{i-1} (X_{i-2} - X_{i+1}) - X_i - \frac{hc}{b} \sum_{j=1}^J Y_{j,i} + F, \label{eq:l96slow}\\
    \frac{d Y_{j,i}}{dt} &= -cbY_{j+1,i} (Y_{j+2,i} - Y_{j-1,i}) - c Y_{j,i} + \frac{hc}{b} X_{i} \label{eq:l96fast},
\end{align}
where Equation~\ref{eq:l96slow} represents the evolution of slow, high-amplitude variables $X_i~(i=1,\dots,n)$, and Equation~\ref{eq:l96fast} provides the evolution of a coupled fast, low-amplitude variable $Y_{j,i}~(j=1,\dots,J)$. We use $n=36$ and $J=10$ in our computational experiments. We utilize $c=10$ and $b=10$, which implies that the small scales fluctuate 10 times faster than the larger scales. Also, the coupling coefficient $h$ between two scales is equal to 1 and the forcing is set at $F=10$ to make both variables exhibit the chaotic behavior. \textcolor{rev1}{The boundary conditions for the slow and fast variables are detained in Section~\ref{sec:results} along with the generation of initial condition for the two-level Lorenz 96 system.}

In parameterization research, small scale variables are not resolved and their effect is typically parameterized as a function of resolved large scale variables. 
A forecast model for the resolved variables given in Equation~\ref{eq:l96slow} can be constructed with the parameterization for unresolved variables as follows 
\begin{align}
    \frac{d \widetilde{X}_i}{dt} &= -\widetilde{X}_{i-1} (\widetilde{X}_{i-2} - \widetilde{X}_{i+1}) - \widetilde{X}_i - \frac{hc}{b} G_i + F, \label{eq:l96slow_p}
\end{align}
where the tilde is used to denote the fact that the parameterization $G_i$ is used to represent the effect of unresolved variables. Typically, the parameterizations is a function of resolved variables and can be written mathematically as 
\begin{equation}
    \sum_{j=1}^J Y_{j,i} :\approx G_i = \mathbf{N}(\mathbf{\widetilde{X}}), \label{eq:parm}
\end{equation}
where $\mathbf{N}(\cdot)$ is the nonlinear mapping of resolved variables to the parameterizations at the $i^{\text{th}}$ grid point. This mapping can be based on certain physical arguments or can also be learned with any data-driven methods. Therefore in parameterization research for multiscale systems, the underlying physical laws governing the dynamics of resolved variables are assumed to be known exactly, and the effect of unresolved variables is considered through parameterizations $G_i$. If we use data-driven methods to represent the parameterization $G_i$, then the forecast model given in Equation~\ref{eq:l96slow_p} can be considered as a hybrid model. Our main objective in this work is to improve the forecasting capability of multiscale systems that are represented by a hybrid model embedded with data-driven parameterizations and we achieve this through data assimilation techniques.   

\textcolor{rev1}{
\subsection{Kraichnan turbulence}
Here, we summarize the mathematical background of subgrid-scale parameterizations in the LES of two-dimensional turbulence. Even though two-dimensional turbulence cannot be realized in practice, it is extensively used for modeling geophysical flows in the atmosphere and ocean \cite{bouchet2012statistical,boffetta2012two}. The confinement of fluid turbulence to two spatial dimensions leads to Kraichnan–Batchelor–Leith (KBL) theory of dual cascade with an inverse energy cascade to larger scales and direct enstrophy cascade to smaller scales. The non-dimensional vorticity transport equation for incompressible flows can be written as 
\begin{align}
     {\frac{\partial \omega}{\partial t}+J(\omega,\psi) = \frac{1}{\text{Re}}\nabla^2\omega}, \\
    {J(\omega, \psi) = \frac{\partial \omega}{\partial x} \frac{\partial \psi}{\partial y} - \frac{\partial \omega}{\partial y} \frac{\partial \psi}{\partial x},}
\end{align}
where $\omega$ is the vorticity, $\psi$ is the stremfunction, $J$ is the Jacobian (or the nonlinear term), and Re is the Reynolds number of the flow. The vorticity and streamfunction are related to each other through the Poisson equation given by
\begin{equation}
    \nabla^2 \psi = -\omega.
\end{equation}
The governing equations for LES are obtained by applying a low-pass filtering operation, and the filtered vorticity transport equation can be written as 
\begin{equation}
    \frac{\partial \overbar{\omega}}{\partial t} + \overbar{J({\omega},{\psi})} = \frac{1}{\text{Re}}\nabla^2\overbar{\omega}.
\end{equation}
The above equation can be rewritten as 
\begin{equation}
    \frac{\partial \overbar{\omega}}{\partial t} + {J(\overbar{\omega},\overbar{\psi})} = \frac{1}{\text{Re}}\nabla^2\overbar{\omega} + \Pi,
\end{equation}
where the overbar quantities represent filtered variables and are evolved on a grid which is significantly coarse than required for the DNS. The effect of the unresolved scales due to truncation of high wavenumber flow scales is encompassed in a subgrid-scale source term $\Pi$ and must be modeled. Mathematically, the true source term $\Pi$ can be expressed as 
\begin{equation}
    {\Pi} = J(\overbar{\omega},\overbar{\psi}) - \overbar{J(\omega, \psi)}.
\end{equation}
The approximation of subgrid processes plays an important role in determining the accuracy of large-scale flows and therefore the subgrid-scale parameterizations are critical to accurate LES simulations of geophysical flows \cite{frederiksen2013subgrid}. Different models have been proposed in the literature for subgrid-scale parameterizations in geophysical flows \cite{smagorinsky1963general,leith1971atmospheric,frederiksen2006dynamical,eden2008towards,san2013approximate,maulik2017stable}, and remains an active area of research due to complexity of the subgrid-scale closure modeling. In the present work, we put forth a data-driven framework based on a neural network to predict the approximate value of the source term $\Pi$ as a function of resolved flow variables on the coarser grid. One of the main advantages of data-driven closure modeling is that they are computationally faster than dynamic closure modeling procedure that involves several test filtering operations \cite{sarghini2003neural,pawar2020apriori,pal2019deep}.            
}
    
\section{Neural Network Parameterizations} \label{sec:nnp}
The parameterization problem in multiscale flows can be posed as a regression problem where the mapping between resolved scales and unresolved scales has to be determined.
We consider supervised class of machine learning algorithms, where the optimal map between inputs and outputs is learned. In this section, we describe an artificial neural network (ANN) also called as multilayer perceptron, and convolutional neural network (CNN) to build data-driven parameterization models. 

\subsection{Artificial neural network} \label{sec:ann}
An artificial neural network is made up of several layers consisting of the predefined number of neurons. Each neuron consists of certain coefficients called weights and some bias. The weight determines how significant certain input feature is to the output. The input from the previous layer is multiplied by a weight matrix as shown below
\begin{equation}
    S^l = \mathbf{W}^l \mathcal{X}^{l-1},
\end{equation}
where $\mathcal{X}^{l-1}$ is the output of the $(l-1)^{\text{th}}$ layer, $\mathbf{W}^l$ is the matrix of weights for the $l^{\text{th}}$ layer. The summation of the above input-weight product and the bias is then passed through a node's activation function which is usually some nonlinear function. The introduction of nonlinearity through activation function allows the neural network to learn highly complex relations between the input and output. The output of the $l^{\text{th}}$ layer can be written as
\begin{equation}
    \mathcal{X}^l = \zeta(S^l+B^l),
\end{equation}
where $B^l$ is the vector of biasing parameters for the $l^{\text{th}}$ layer and $\zeta$ is the activation function. If there are $L$ layers between the input and the output in a neural network, then the output of the neural network can be represented mathematically as follows
\begin{equation}
    \tilde{\mathcal{Y}} =\zeta_L(\mathbf{W}^L,B^L,\dots,\zeta_2(\mathbf{W}^2,B^2,\zeta_1(\mathbf{W}^1,B^1,\mathcal{X}))),
\end{equation}
where $\mathcal{X}$ and $\mathcal{\tilde{Y}}$ are the input and output of the ANN, respectively. There are several activation functions that provides different nonlinearity. Some of the widely used activation functions are sigmoid $\zeta(\phi) = 1/(1+e^{-\phi})$, hyperbolic tangent (tanh) $\zeta(\phi) = (e^{\phi}-e^{-\phi})/(e^{\phi}+e^{-\phi})$, and rectified linear unit (ReLU) $\zeta(\phi) = \text{max}[0,\phi]$.

The matrix $\mathbf{W}$ and $B$ are determined through the minimization of the loss function (for example mean squared error between true and predicted labels). The gradient of the objective function with respect to weights and biases are calculated with the backpropagation algorithm. The optimization algorithms like the stochastic gradient descent method \cite{kingma2014adam} provide a rapid way to learn optimal weights. The training procedure for ANN can be summarized as:
\begin{itemize}
    \item The input and output of the neural network are specified along with some initial weights initialization for neurons.
    \item The training data is run through the network to produce output $\tilde{\mathcal{Y}}$ whose true label is $\mathcal{Y}$.
    \item The derivative of the objective function with each of the training weight is computed using the chain rule.
    \item The weights are then updated based on the learning rate and the optimization algorithm.
\end{itemize}
We continue to iterate through this procedure until convergence or the maximum number of iterations is reached. There are different ways in which the relationship between resolved and unresolved variables in multiscale systems can be learned with the ANN. The most common method is to employ point-to-point mapping, where the input features at a single grid point are utilized to learn the output labels at that point \cite{xie2019artificial,yang2019predictive,gamahara2017searching}. Another method is to include the information at neighboring grid points to determine the output label at a single point \cite{maulik2019subgrid,pawar2020apriori}. For a two-level Lorenz system, we train our ANN by including information at different number of neighboring grid points and assess how does this additional information affects in learning the correlation between resolved and unresolved variables. We investigate three types of ANN models and they can be written as 
\begin{align} 
    \text{ANN-3} : \{X_{i-1},X_i,X_{i+1}\} \in \mathbb{R}^3 &\rightarrow \{ \tilde{G_i} \} \in \mathbb{R}^1, \label{eq:ann3} \\
    \text{ANN-5} : \{X_{i-2}\dots,X_{i+2}\} \in \mathbb{R}^5 &\rightarrow \{ \tilde{G_i} \} \in \mathbb{R}^1, \label{eq:ann5} \\
    \text{ANN-7} : \{X_{i-3}\dots,X_{i+3}\} \in \mathbb{R}^7 &\rightarrow \{ \tilde{G_i} \} \in \mathbb{R}^1, \label{eq:ann7}
\end{align}
where $\tilde{G_i}$ is the predicted parameterization at $i^{\text{th}}$ grid point and $X_{i}$ is the resolved variable. For the training, we assume that the resolved variables and the parameterizations are known exactly and are computed by solving Equation~\ref{eq:l96slow} and Equation~\ref{eq:l96fast} in a coupled manner. For all ANN architectures used in this study, we apply two hidden layers with 40 neurons and ReLU activation function \textcolor{rev1}{for all hidden layers. For the output layer, the linear activation function is used.} The ANN is trained using an Adam optimizer for 300 iterations.   

\subsection{Convolutional neural network} \label{sec:cnn}
The convolutional neural network (CNN) is particularly attractive when the data is in the form of two-dimensional images \cite{lecun1998gradient}. Here, we present the CNN architecture assuming that the input and output of the neural network have the structure of two-dimensional images. This formulation can be easily applied to one-dimensional images when the dimension in one direction is collapsed to one. The Conv layers are the fundamental building blocks of the CNN. Each Conv layer has a predefined number of filters (also called kernels) whose weights have to be learned using the backpropagation algorithm. The shape of the filter is usually smaller than the actual image and it extends through the full depth of the input volume from the previous layer. 
For example, if the input to the CNN has $256 \times 256 \times 1$ dimension where 1 is the number of input features, the kernels of the first Conv layer can have $3 \times 3 \times 1$ shape. During the forward propagation, the filter is convolved across the width and height of the input volume to produce the two-dimensional map. The two-dimensional map is constructed by computing the dot product between the weights of the filter and the input volume at any position and then sliding it over the whole volume. Mathematically the convolution operation corresponding to one filter can be written as 
\begin{equation}
    S^l_{ij} = \sum_{p=-\Delta_i/2}^{\Delta_i/2} \sum_{q=-\Delta_j/2}^{\Delta_j/2} \sum_{r=-\Delta_k/2}^{\Delta_k/2} \mathbf{W}_{pqr}^l \mathcal{X}_{i+p ~ j+q ~ k+r}^{l-1} + B_{pqr},
\end{equation}
where $\Delta_i$, $\Delta_j$, $\Delta_k$ are the sizes of filter in each direction, $\mathbf{W}_{pqr}^l$ are the entries of the filter for $l^{\text{th}}$ Conv layer, $ B_{pqr}$ is the biasing parameter, and $\mathcal{X}_{ijk}^{l-1}$ is the input from $(l-1)^{\text{th}}$ layer. Each Conv layer will have a set of predefined filters and the two-dimensional map produced by each filter is then stacked in the depth dimension to produce a three-dimensional output volume. This output volume is passed through an activation function to produce a nonlinear map between inputs and outputs. The output of the $l^{\text{th}}$ layer is given by 
\begin{equation}
    \mathcal{X}_{ijk}^l = \zeta(S_{ijk}^l),
\end{equation}
where $\zeta$ is the activation function. It should be noted that as we convolve the filter across the input volume, the size of the input volume shrinks in height and width dimension. Therefore, it is common practice to pad the input volume with zeros called zero-padding. The zero-padding permits us to control the shape of the output volume and is used in our neural network parameterization framework to preserve the shape so that input and output width and height are the same. The main advantage of CNN is its weight sharing property because the filter of the smaller size is shared across the whole image which is larger in size. This allows CNN to handle large data without the significant computational overhead. The CNN mapping for learning parameterizations in a two-level Lorenz model can be mathematically presented as 
\begin{align} 
    \text{CNN} : \{X_{1},\dots,X_{n}\} \in \mathbb{R}^n &\rightarrow \{ \tilde{G_1}, \dots \tilde{G_{n}} \} \in \mathbb{R}^n, \label{eq:cnn}
\end{align}
where $X_i$ is the resolved variable and $\tilde{G_i}$ is the predicted parameterization. Therefore, the solution at a single time step corresponds to one training example for training the CNN. In our CNN architecture, we use only one hidden layer between the input and output. This hidden layer has 128 filters with $7 \times 1$ shape. We apply ReLU activation function \textcolor{rev1}{for all hidden layers and the linear activation function for the output layer.} Also, the zero-padding is used to keep the input and output shape the same. The CNN is trained with an Adam optimizer for 400 iterations. \textcolor{rev1}{The hyperparameters of both ANN and CNN architectures were obtained through parametric study for a different number of neurons/filters and the number of hidden layers. 80\% of the total data selected randomly was used for training and the remaining 20\% of the data was used for validation. The selection of hyperparameters is done in such a way that the mean squared error between the actual and predicted parameterization drops smoothly for both training and validation dataset so that overfitting is avoided and our model generalizes well to the unseen data. There are other methods like regularization, dropout, ensembling from different models, early stopping that can be adopted to prevent overfitting. An extensive hyperparameter search can be carried out for complex geophysical flows using neural architecture search packages like DeepHyper\cite{maulik2020recurrent} and Tune\cite{liaw2018tune}. }

\textcolor{rev1}{
For the two-dimensional turbulence, we learn the source term $\Pi$ as a function of resolved flow variables, i.e., the vorticity $\omega$, the streamfunction $\psi$, and two eddy-viscosity kernels as input features. The use of eddy-viscosity kernels as input features can be considered as a feature engineering step where certain important quantities are pre-computed and the neural network is presented with it as raw input features so as to facilitate the faster training and robust prediction. The CNN map for two-dimensional turbulence can be mathematically written as    
\begin{align}
    \text{CNN} : \{\bar{\omega}, \bar{\psi}, |\bar{S}|, |\nabla \bar{\omega }|\}  &\rightarrow \{ \tilde{\Pi} \},
\end{align}
where $\tilde{\Pi}$ is the predicted source term, $|\bar{S}|$ is the Smagorinsky kernel, and $|\nabla \bar{\omega }|$ is the Leith kernel. The Smagorinsky and Leith kernels are computed as follow
\begin{align}
    |\bar{S}| &= \sqrt{4\bigg(\frac{{\partial^2\bar{\psi}}}{\partial x \partial y} \bigg)^2 + \bigg( \frac{{\partial^2\bar{\psi}}}{\partial x^2}  - \frac{{\partial^2\bar{\psi}}}{\partial y^2}\bigg)^2}, \\
    |\nabla \bar{\omega }| &= \sqrt{\bigg(\frac{\partial \bar{\omega}}{\partial x} \bigg)^2 + \bigg(\frac{\partial \bar{\omega}}{\partial y} \bigg)^2}.
\end{align}
The CNN architecture to learn the parameterization model for Kraichnan turbulence consist of 6 hidden layers and 16 filters in each hidden layers. The  size of the filter in each hidden layer is $3 \times 3$ and the ReLU activation function is utilized for hidden layers. The training is performed for 800 epochs using the Adam optimizer. We note here that the predicted source term by the CNN is further post-processed during the deployment before it is injected into the vorticity transport equation to ensure numerical stability and we detail that procedure in Section~\ref{sec:results}.
}

\section{Data assimilation } \label{sec:da}
As highlighted in many studies, neural network parameterizations suffer from instabilities and biases once the trained model is deployed in a forward solver \cite{rasp2018deep,rasp2018deep,brenowitz2018prognostic,brenowitz2019spatially,wu2019reynolds}. From our numerical experiments with the two-level Lorenz system, we observe that the forward model with only neural network parameterizations delivers accurate prediction only up to some time and after that the model starts deviating from the true trajectory. In order to address this issue and improve the long-term forecast with hybrid models, we utilize the data assimilation (DA) to incorporate noisy measurements for the prediction of future state. The main theme of DA is to extract the information from observational data to correct dynamical models and improve their prediction. There is a rich literature on DA\cite{lewis2006dynamic,simon2006optimal,evensen2009data,gelb1974applied,welch1995introduction} and here we discuss only sequential data assimilation problem and then outline the algorithm procedure for \textcolor{rev1}{perturbed observations ensemble Kalman filter (EnKF)}, and the deterministic ensemble Kalman filter (DEnKF).

We consider the dynamical system whose evolution can be represented as 
\begin{equation}\label{eq:dyn_model}
    \mathbf{x}_{k+1} = \textbf{M}_{t_k \rightarrow t_{k+1}}(\mathbf{x}_{k}) + \mathbf{w}_{k+1},
\end{equation}
where $\mathbf{x}_k \in \mathbb{R}^n$ is the state of the dynamical system at discrete time $t_k$, $\textbf{M}:\mathbb{R}^n \rightarrow \mathbb{R}^n$ is the nonlinear model operator that defines the evolution of the system. 
The term $\mathbf{w}_{k+1}$ denotes the model noise that takes into account any type of uncertainty in the model that can be attributed to boundary conditions, imperfect models, etc. Let $\mathbf{z}_k \in \mathbb{R}^m$ be observations of the state vector obtained from noisy measurements and cane be written as
\begin{equation}
    \mathbf{z}_k = \mathbf{h}(\mathbf{x}_k) + \mathbf{v}_k,
\end{equation}
where $\mathbf{h}(\cdot)$ is a nonlinear function that maps $\mathbb{R}^n \rightarrow \mathbb{R}^m$, and $\mathbf{v}_k \in \mathbb{R}^m$ is the measurement noise. We assume that the measurement noise is a white Gaussian noise with zero mean and the covariance matrix $\mathbf{R}_k$, i.e., $\mathbf{v}_k \sim {\cal{N}}(0,\mathbf{R}_k)$. Additionally, the noise vectors $\mathbf{w}_{k}$ and $\mathbf{v}_{k}$ are assumed to be uncorrelated to each other at all time steps. The sequential data assimilation can be considered as a problem of estimating the state $\mathbf{x}_k$ of the system given the observations up to time $t_k$, i.e., $\mathbf{z}_1,\dots,\mathbf{z}_k$. When  we utilize observations to estimate the state of the system, we say that the data are assimilated into the model. We will use the notation $\widehat{\mathbf{x}}_k$ to denote an analyzed state of the system at time $t_k$ when all of the observations up to and including time $t_k$ are used in determining the state of the system. When all the observations before (but not including) time $t_k$ are utilized for estimating the state of the system, then we call it the forecast estimate and denote it as $\mathbf{x}^f_k$.

\textcolor{rev1}{The ensemble Kalman filter (EnKF) \cite{burgers1998analysis} follows the Monte Carlo approach to approximate the probability distribution in the Kalman filter equations \cite{kalman1960new}. We start by initializing the state of the system for different ensemble members as follows   
\begin{align}
    \mathbf{\widehat{X}}_0(i) &= \mathbf{m}_0 + \mathbf{y}_0(i), \quad i=1 \dots N \label{eq:enkf_xo}
\end{align}
where $\mathbf{y}_0(i) \sim {\cal{N}} (0,\mathbf{P}_0)$, $\mathbf{m}_0$ is some assumed mean state of the system, $\mathbf{P}_0$ is the initial covariance error matrix, and $N$ is the number of ensemble members. The propagation of the state for each ensemble over the time interval $[t_k,t_{k+1}]$ can be written as
\begin{align}
\mathbf{X}^f_{k+1}(i) &= \mathbf{M}_{t_k \rightarrow t_{k+1}}(\widehat{\mathbf{X}}_k(i)) + \mathbf{w}_{k+1}.
\label{eq:enkf_xk}
\end{align}
The term $\mathbf{w}_{k+1}$ accounting for model imperfections is usually assumed to be Gaussian noise. In this study, we consider the model error by means of multiplicative inflation \cite{anderson1999monte} and without loss of generality we set $\mathbf{w}_{k+1} = 0$. The prior state and the prior covariance matrix are approximated using the sample mean and error covariance matrix $\mathbf{P}^f_{k+1}$ as follows
\begin{align}
    \mathbf{x}^f_{k+1} &= \frac{1}{N} \sum_{i=1}^N \mathbf{X}^f_{k+1}(i), \label{eq:enkf_xf}\\
\mathbf{E}^f_{k+1}(i) &= \mathbf{X}^f_{k+1}(i) - \mathbf{x}^f_{k+1}, \\
\mathbf{P}^f_{k+1} &= \frac{1}{N-1} \sum_{i=1}^N  \mathbf{E}^f_{k+1}(i) (\mathbf{E}^f_{k+1}(i))^{\text{T}}. \label{eq:enkf_pf}
\end{align}
Once the observations are available at time $t_{k+1}$, we generate $N$ realizations of perturbed observations as follows
\begin{align}
    {\mathbf{Z}}_{k+1}(i) &= {\mathbf{z}}_{k+1} + \mathbf{v}_{k+1}(i), \label{eq:enkf_zv}
\end{align} 
where $\mathbf{v}_{k+1}(i) \sim {\cal{N}} (0,\mathbf{R}_{k+1})$. Each member of the forecast ensemble $\mathbf{X}^f_{k+1}(i)$ is analyzed using the Kalman filter formulae as shown below 
\begin{align}
    \widehat{\mathbf{X}}_{k+1}(i) &= \mathbf{X}^f_{k+1}(i) + \mathbf{K}_{k+1}[\mathbf{Z}_{k+1}(i) - \mathbf{h} (\mathbf{X}^f_{k+1}(i))], \label{eq:enkf_xe} \\
    \mathbf{K}_{k+1} &= \mathbf{P}^f_{k+1} \mathbf{H}_{k+1}^\text{T}[ \mathbf{H}_{k+1} \mathbf{P}^f_{k+1} \mathbf{H}_{k+1}^\text{T} + \mathbf{R}_{k+1}]^{-1},
\end{align}
where $\mathbf{H} \in \mathbb{R}^{m \times n}$ is the Jacobian of observation function $\mathbf{h}(\cdot)$. The analysis state estimate at time $t_{k+1}$ is computed using the sample mean of all ensemble members as 
\begin{align}
    \widehat{\mathbf{x}}_{k+1} &= \frac{1}{N} \sum_{i=1}^N \widehat{\mathbf{X}}_{k+1}(i). \label{eq:enkf_xa}
\end{align}
In order to take model imperfections into account, all ensemble members are updated by applying inflation to all ensemble anomalies as follows
\begin{align}
    \widehat{\mathbf{X}}_{k+1}(i) \leftarrow \widehat{\mathbf{x}}_{k+1} + \lambda \cdot (\widehat{\mathbf{X}}_{k+1}(i) - \widehat{\mathbf{x}}_{k+1} ), \label{eq:enkf_inflation}
\end{align}
where $\lambda$ is the inflation factor. The inflation also helps to address the problem of covariance underestimation due to small number of ensembles \cite{evensen2009data}. The inflation factor can either be a scalar or it can be made space and time dependent to improve the filter performance \cite{anderson2007adaptive,attia2018optimal}. In this study, we use the constant value of the inflation factor over the entire space at all times. 
}

As a variant of the low-rank sequential nonlinear filtering framework, we also utilize the deterministic EnKF (DEnKF) algorithm proposed by \citet{sakov2008deterministic} for the data assimilation. We start the DEnKF algorithm by initializing the state estimate for all ensemble members similar to the EnKF algorithm as given in Equation~\ref{eq:enkf_xo}. The anomalies between the forecast estimate of all ensembles and its sample mean (calculated using Equation~\ref{eq:enkf_xf}) is 
\begin{align}
    \mathbf{A}^f_{k+1}(i) &= \mathbf{X}^f_{k+1}(i) - \mathbf{x}^f_{k+1}. \label{eq:denkf_an}
\end{align}
Once the observations are available at time $t_{k+1}$, the forecast state estimate is assimilated using the Kalman filter analysis equation as follows 
\begin{align}
    \widehat{\mathbf{x}}_{k+1} = \mathbf{x}^f_{k+1} + \mathbf{K}_{k+1}[\mathbf{z}_{k+1} - \mathbf{h}(\mathbf{x}^f_{k+1})]. \label{eq:denkf_xa}
\end{align}
Here, the Kalman gain matrix is computed using its square root version (without storing or computing $\mathbf{P}^f_{k+1}$ explicitly) as follows
\begin{multline}
    \mathbf{K}_{k+1} = \frac{\mathcal{A}_{k+1}^f(\mathbf{H}_{k+1} \mathcal{A}_{k+1}^f)^{\text{T}}}{N-1} \bigg[\frac{(\mathbf{H}_{k+1}\mathcal{A}_{k+1}^f)(\mathbf{H}_{k+1}\mathcal{A}_{k+1}^f)^\text{T}}{N-1}\\
    + \mathbf{R}_{k+1} \bigg]^{-1},
\end{multline}
where $\mathbf{H} \in \mathbb{R}^{m \times n}$ is the Jacobian of the observation operator (i.e., $H_{kl} = \frac{\partial h_k}{ \partial x_l}$), and the matrix $\mathcal{A}_{k+1}^f \in \mathbb{R}^{n \times N}$ is concatenated as follows
\begin{align}
\mathcal{A}_{k+1}^f = [\mathbf{A}^f_{k+1}(1), \mathbf{A}^f_{k+1}(2), \dots, \mathbf{A}^f_{k+1}(N)].
\end{align}
The anomalies for all ensemble members are then updated separately with half the Kalman gain as shown below 
\begin{align}
    \widehat{\mathbf{A}}_{k+1}(i) = \mathbf{A}^f_{k+1}(i) - \frac{1}{2}\mathbf{K}_{k+1} \mathbf{H}_{k+1}\mathbf{A}^f_{k+1}(i). \label{eq:denkf_aa}
\end{align}
The analysis state for all ensemble members is obtained by adding ensemble anomalies and can be written as
\begin{align}
    \widehat{\mathbf{X}}_{k+1}(i) = \widehat{\mathbf{x}}_{k+1} + \lambda \cdot \widehat{\mathbf{A}}_{k+1}(i), \label{eq:denkf_ea}
\end{align}
\textcolor{rev1}{where $\lambda$ is the inflation factor.} We validate our implementation of the DEnKF algorithm using the one-level Lorenz 96 model and is detailed in Appendix~\ref{app:validation}.

\section{Numerical Experiments}\label{sec:results}
In the following, we present the findings of our numerical experiments with a two-level Lorenz 96 model and Kraichnan turbulence. 

\subsection{Two-level Lorenz 96 model}
In this subsection, we discuss the results of numerical experiments with a two-level variant of the Lorenz 96 system embedded with neural network parameterizations for the unresolved variables. We utilize the fourth-order Runge-Kutta numerical scheme with a time step $\Delta t = 0.001$ for temporal integration of the Lorenz 96 model. We apply the periodic boundary condition for the slow variables, i.e., $X_{i-n}=X_{i+n}=X_i$. The fast variables are extended by letting $Y_{j,i-n}=Y_{j,i+n}=Y_{j,i}$, $Y_{j-J,i}=Y_{j,i-1}$, and $Y_{j+J,i}=Y_{j,i+1}$.
The physical initial condition is computed by starting with an equilibrium condition at time $t=-5$ for slow variables. The equilibrium condition for slow variables is $X_i=F$ for $i \in {1,2,\dots,n}$. We perturb the equilibrium solution for the $18^{\text{th}}$ state variable as $X_{18}=F+0.01$. At the time $t=-5$, the fast variables are assigned with random numbers between $-F/10$ to $F/10$. We integrate a two-level Lorenz 96 model by solving both Equation~\ref{eq:l96slow} and Equation~\ref{eq:l96fast} in a coupled manner up to time $t=0$. With this initial condition (i.e., at $t=0$), we generate the training data for neural networks by integrating the two-level Lorenz 96 model from $t=0$ to $t=10$. Therefore, we gather 10,000 temporal snapshots to generate the training data. For all our numerical experiments, we use 80\% of the data to train the neural network and 20\% data to validate the training. We assess the performance of a trained neural network by deploying it in a forecast model for temporal integration between time $t=10$ to $t=20$. Therefore, there is no overlap between the data used for training and testing. Since the neural network has not seen the testing data during the training, the performance of neural network parameterizations in this temporal region will give us an insight on its generalizability to unseen data.   

\begin{figure*}[htbp]
\centering
\mbox{\subfigure{\includegraphics[width=0.8\textwidth]{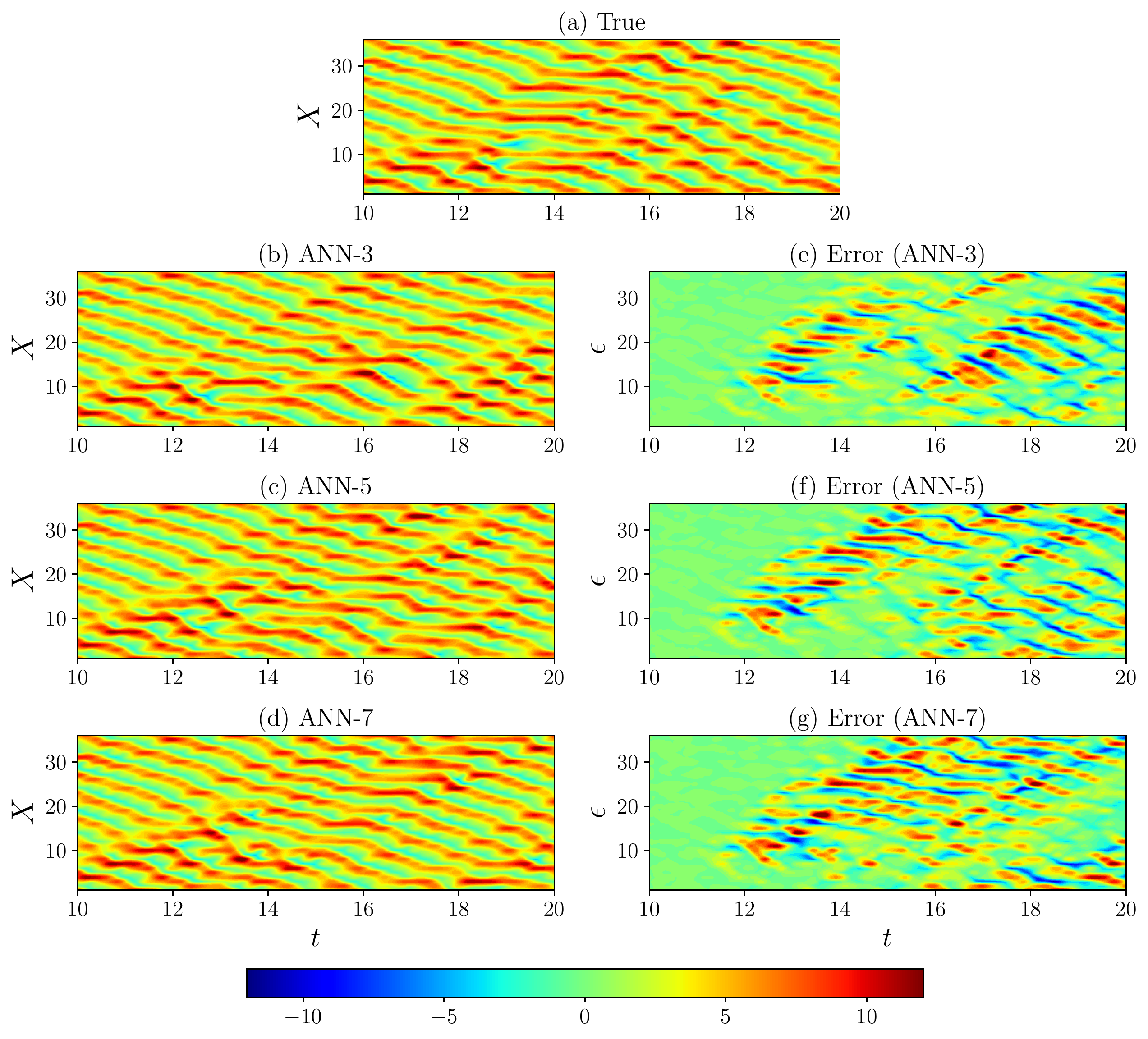}}}
\caption{Full state trajectory of the multiscale Lorenz 96 model with the closure term computed using the different neighboring stencil mapping feedforward ANN architecture.}
\label{fig:ml_ann}
\end{figure*}

First, we present results for ANN based parameterizations trained using neighboring stencil mapping as discussed in Section~\ref{sec:ann}. Figure~\ref{fig:ml_ann} displays the full state trajectory of the Lorenz 96 model from time $t=10$ to $t=20$ computed by solving both the evolution of slow and fast variables (i.e., True) and with ANN based parameterizations for fast variables (i.e., ANN-3, ANN-5, ANN-7). The difference between the true solution field and the predicted solution field is also depicted in Figure~\ref{fig:ml_ann}. It can be observed that the predicted solution field starts deviating from the true solution field at around $t \approx 12$ for all ANN-based parameterizations.

\begin{figure*}[htbp]
\centering
\mbox{\subfigure{\includegraphics[width=0.8\textwidth]{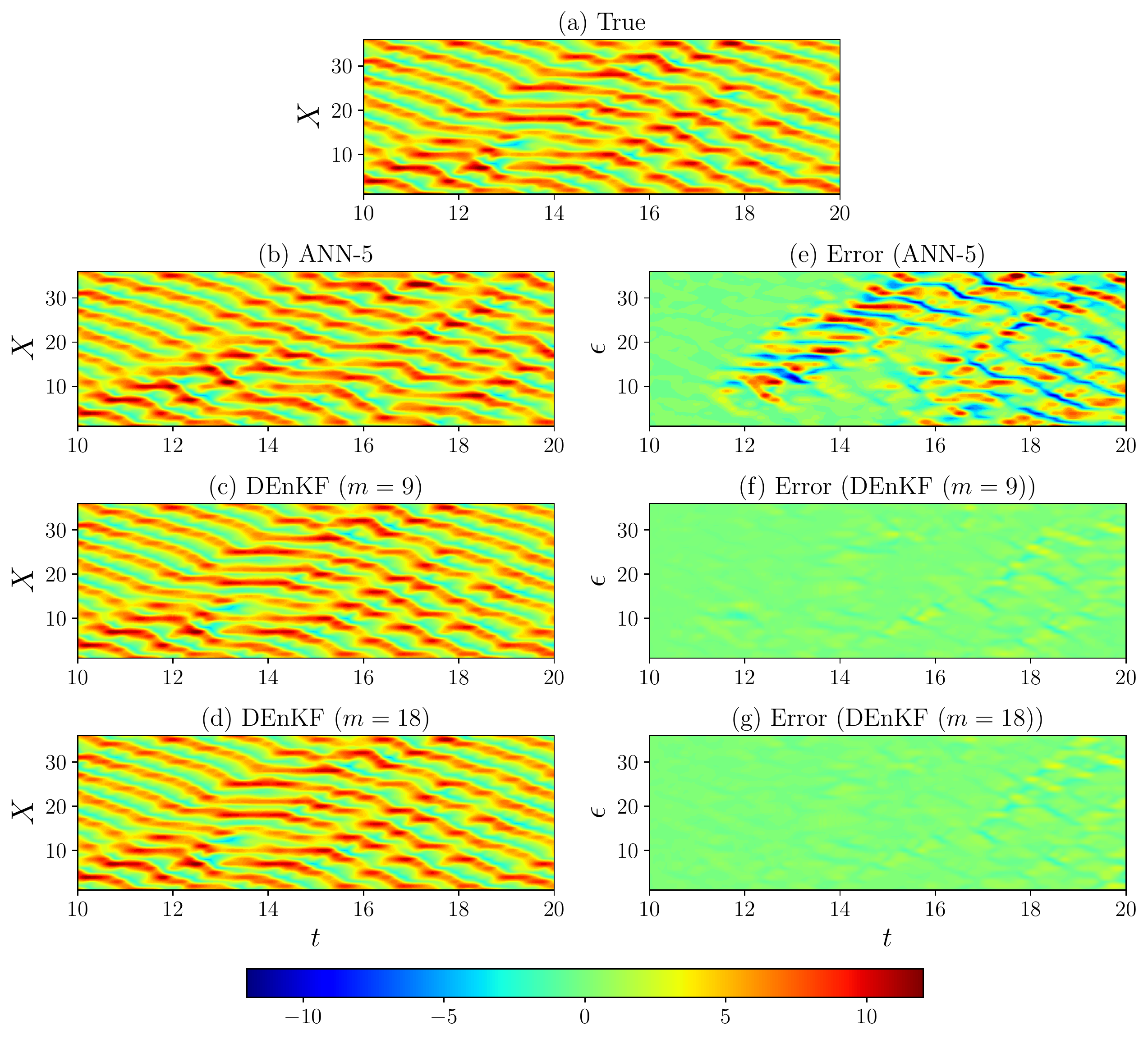}}}
\caption{Full state trajectory of the multiscale Lorenz 96 model with the closure term computed using the five-point neighboring stencil mapping feedforward ANN architecture and the DEnKF used for data assimilation.} 
\label{fig:ml_ann_da}
\end{figure*}

Next, we illustrate how the prediction of a two-level Lorenz 96 model with neural network parameterizations can be improved using sequential data assimilation by incorporating noisy observations in the future state prediction. For our twin experiment, we obtain observations by adding noise drawn from the Gaussian distribution with zero mean and the covariance matrix $\mathbf{R}_k$, i.e., $\mathbf{v}_k \sim {\cal{N}}(0,\mathbf{R}_k)$. We use $\mathbf{R}_k = \sigma_b^2 \mathbf{I}$, where $\sigma_b$ is the standard deviation of measurement noise and is set at $\sigma_b=1$. We assume that observations are sparse in space and are collected at every $10^{\text{th}}$ time step. The number of ensemble members used for all numerical experiments is $N=30$. We present two levels of observation density in space for the DA. For the first case, we employ observations at $[X_4,X_{8},\dots,X_{36}] \in R^9$ for the assimilation. The second set of observations consists of 50\% of the full state of the system, i.e., $[X_2,X_{4},\dots,X_{36}] \in R^{18}$. In Figure~\ref{fig:ml_ann_da}, we provide the full state trajectory prediction for the ANN-5 parameterization without any DA and with DA for two sets of observations. We can observe that there is a substantial improvement in the long-term prediction even with only 25\% of the observations incorporated through the DEnKF algorithm. The results in Figure~\ref{fig:ml_ann_da} provide the evidence for the good performance of the present framework in achieving accurate long-term prediction for hybrid models embedded with data-driven parameterizations. Therefore, the present framework can lead to accurate forecasting by exploiting online measurements coming from various types of sensor networks and can find applications in different fields like climate modeling, turbulence closure modeling where the subgrid scale parameterizations are unavoidable.  


\begin{figure*}[htbp]
\centering
\mbox{\subfigure{\includegraphics[width=0.8\textwidth]{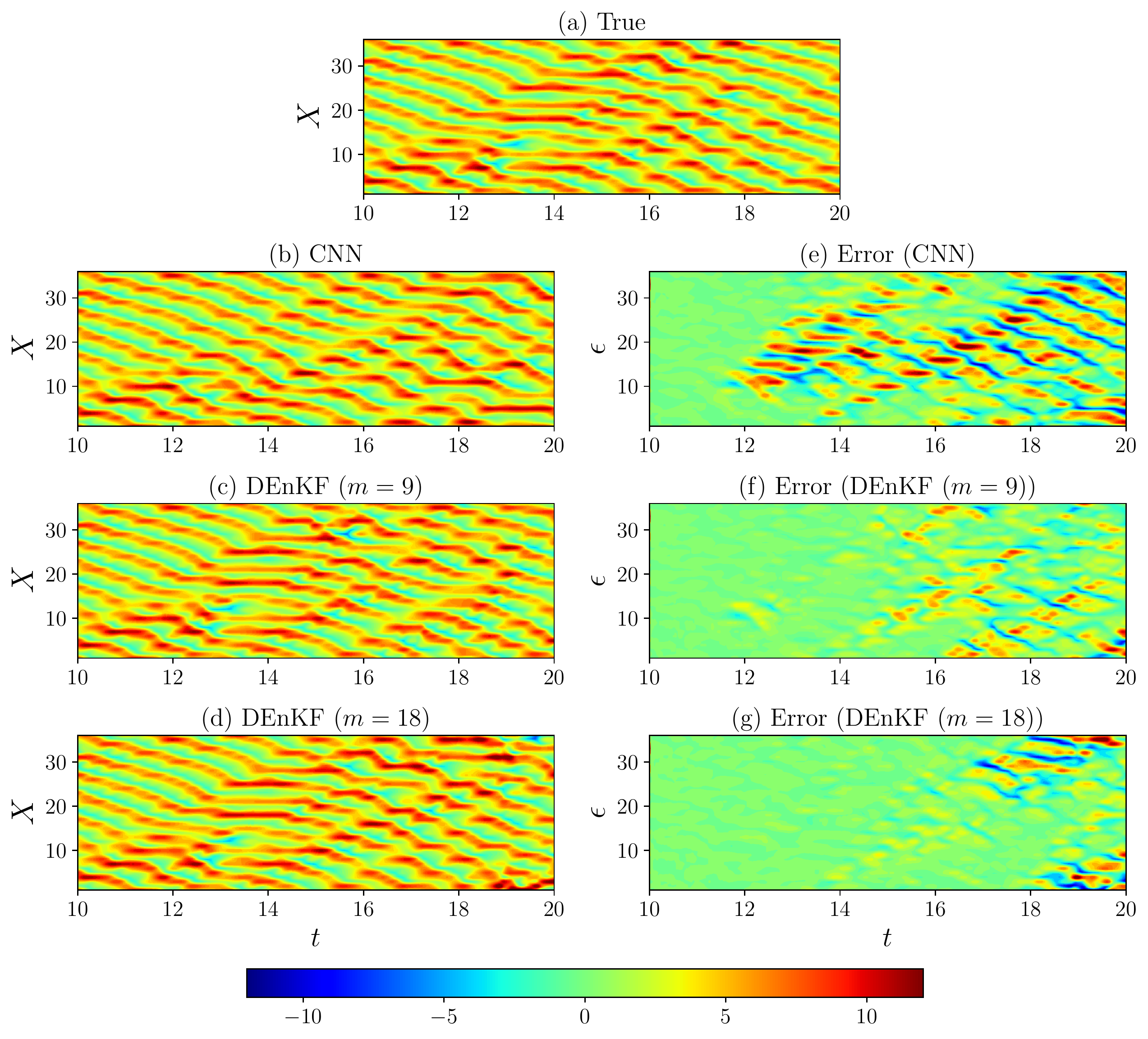}}}
\caption{Full state trajectory of the multiscale Lorenz 96 model with the closure term computed using the CNN architecture and the DEnKF used for data assimilation} 
\label{fig:ml_cnn}
\end{figure*}

Figure~\ref{fig:ml_cnn} illustrates the time evolution of the full state trajectory of a two-level Lorenz 96 model with CNN based parameterizations for unresolved scales. CNN is fed with the entire state of the slow variables as an input and it calculates the parameterizations of fast variables at all grid points. From Figure~\ref{fig:ml_cnn}, we can deduce that the predicted state trajectory starts deviating from the true state at around $t \approx 12$ when only CNN based parameterizations are employed in the forward model of slow variables. When we incorporate observations through DA, we observe considerable improvement in the state prediction over a longer period.  

Based on results presented in Figure~\ref{fig:ml_ann_da} and Figure~\ref{fig:ml_cnn}, we can notice that the error is slightly higher between time $t=18$ to $t=20$ for the CNN based parameterizations empowered with DA. One reason for the inaccurate forecast can be attributed to the uncertainty in the prediction of parameterizations by CNN. \textcolor{rev1}{We highlight here that both ANN-5 and CNN architectures used in this study have similar number of trainable parameters. However, we see a better performance of the ANN-5 architecture over CNN due to a more number of training examples in the case of the ANN. For the ANN, every single point of the two-level Lorenz 96 system is one training example and therefore a single time snapshot of the training data leads to 36 samples for training. However, in the case of CNN, the total number of training samples is equal to the total number of time snapshots available for training. Therefore, we observe the better performance of the ANN-5 over CNN.}

Another potential reason for this discrepancy can be the stochastic nature of the parameterization model. The true parameterization model in itself is stochastic and might not follow a Gaussian distribution. To isolate the source of error, we integrate the forecast model for a two-level Lorenz 96 model without any parameterizations. The two-level Lorenz 96 model with no parameterizations is equivalent to setting the coupling coefficient $h=0$ in Equation~\ref{eq:l96slow} and it reduces to one-level Lorenz 96 model as presented in Equation~\ref{eq:l96}. We note here that the observations used for data assimilation are the same as the numerical experiments with a two-level Lorenz 96 model. Therefore, the effect of unresolved scales is embedded in observations. The parameterization of fast variables (i.e.,$ \frac{hc}{b}\sum_{j=1}^J Y_{j,i}$ term in Equation~\ref{eq:l96slow}) can be considered as an added noise to the true state of the system for a one-level Lorenz 96 model presented in Equation~\ref{eq:l96}. 

\begin{figure*}[htbp]
\centering
\mbox{\subfigure{\includegraphics[width=0.8\textwidth]{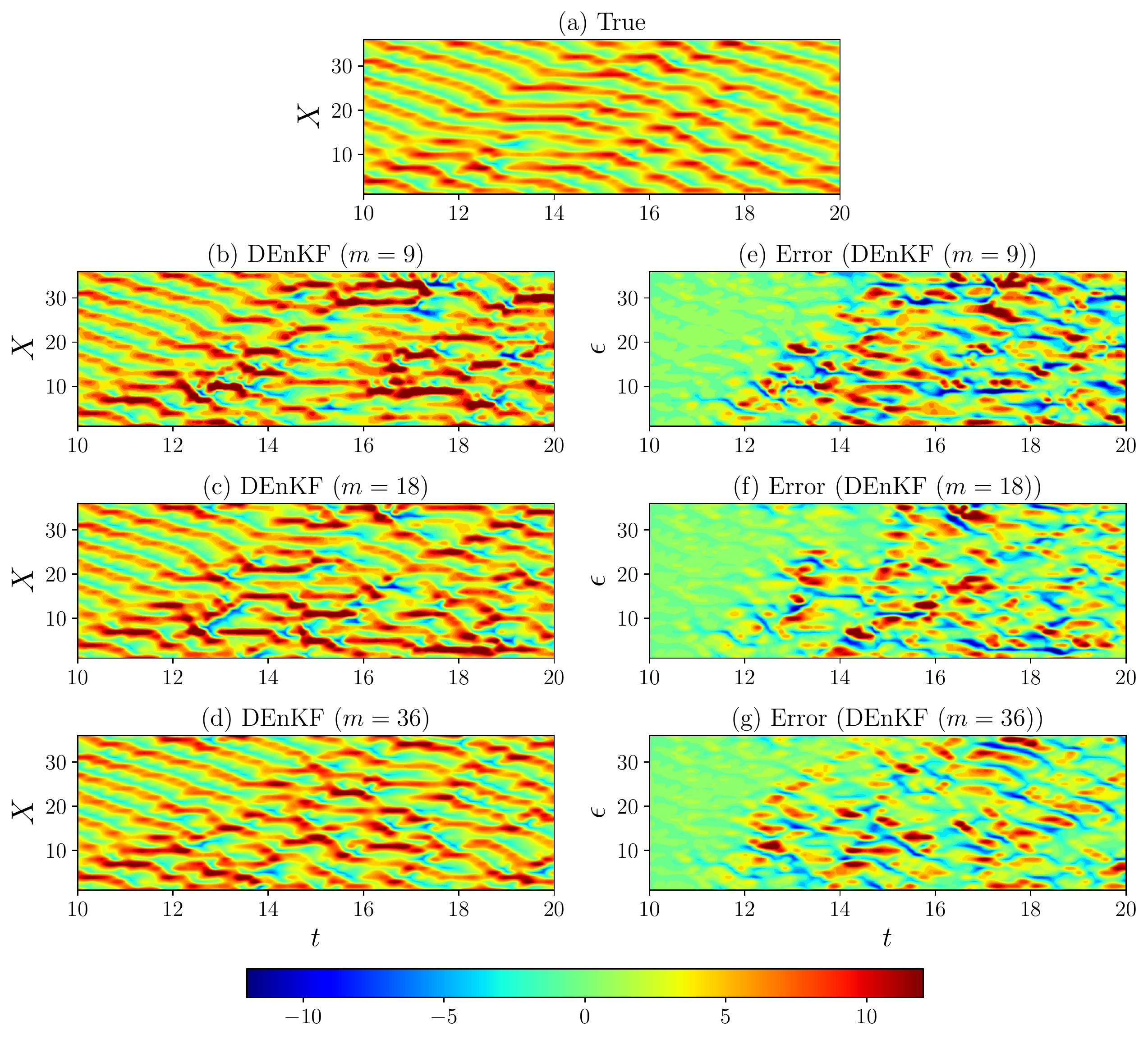}}}
\caption{Full state trajectory of the multiscale Lorenz 96 model with no closure for subgrid processes. The observation data for the DEnKF algorithm is obtained by adding measurement noise to the exact solution of the multiscale Lorenz 96 system.} 
\label{fig:l_nc_da}
\end{figure*}

In Figure~\ref{fig:l_nc_da}, we report the true state of a two-level Lorenz 96 model and also the predicted state trajectory using the DA framework with no parameterization. We provide the results for three sets of observations utilized in DA. The observations are incorporated at every $10^{\text{th}}$ time step of the model through assimilation stage. We can observe that, even when 100\% of the full state is observable, we do not recover the true state trajectory of a two-level Lorenz 96 model. With this observation, we can conclude that it is essential to incorporate parameterization of unresolved scales into a forward model of the DA procedure to recover the accurate state trajectory. The root mean squared error between the assimilated states and true states for three sets of observations is provided in Table~\ref{tab:rmse}.

In this numerical experiment with the truncated model, the observations include the effect of unresolved scales and can be considered as an added noise. The sequential DA methods based on Kalman filters deliver a considerably accurate solution when the model and observations noise is drawn from a Gaussian distribution and enough observations are provided. 
If the parameterization of unresolved scales follows a Gaussian distribution, we should be able to recover the accurate state of the system as the density of observations is increased. However, as reported in Figure~\ref{fig:l_nc_da}, there is a high level of inaccuracy even when 100\% of the state is observable. Therefore, we can conclude that there is a considerable benefit of including neural network parameterizations compared to using no parameterization in the forecast model. The results provided in Figure~\ref{fig:ml_ann_da} and Figure~\ref{fig:ml_cnn} also shows that the neural network parameterizations can capture the non-Gaussian statistics of subgrid scale processes and this leads to accurate forecasting over a longer period. There are other DA approaches that deal with non-Gaussian distributions for noise vectors \cite{li2018trimmed,anderson2010non,apte2007sampling, zupanski2005maximum,carrassi2008maximum,nino2020maximum}. We restrict ourselves to the DEnKF algorithm for DA in this study and plan to explore other DA algorithms in our future work.   

We assess the quantitative performance of different numerical experiments performed in this study using the root mean squared error (RMSE) between the true and predicted state of slow variables in a two-level Lorenz 96 model. The RMSE is computed as shown below 
\begin{equation}\label{eq:rmse}
 \text{RMSE} = \sqrt{\frac{1}{n}\frac{1}{n_t}\sum_{i=1}^{n} \sum_{k=1}^{n_t} \big(X_i^{\text{T}}(t_k) - {X}_i^{\text{P}}(t_k)\big)^2 },
\end{equation}
where $X_i^{\text{T}}$ is the true state of the system and $X_i^{\text{P}}$ is the predicted state of the system. Table~\ref{tab:rmse} reports the RMSE for a two-level Lorenz 96 model for all cases investigated in this work. We can see that the RMSE is very high when we do not use any parameterizations for unresolved scales even when measurements for an entire state of the system are incorporated through DA. The data assimilation alone can not account for the effect of unresolved scales, even though their effect is present in the observations data. Therefore, it is imperative to include parameterizations of fast variables in the forecast model of slow variables. We observe that the ANN architecture provides slightly more accurate results than the CNN based parameterizations for fast variables. 
Also, the RMSE is minimum for the ANN-3 parameterizations and we observe a slight increase in RMSE by including more neighboring information. One potential reason for this observation can be the use of the same hyperparameters for all ANN architectures. However, this change is very small and the RMSE is the same order of magnitude for all types of neural network parameterizations. The RMSE is almost the same when 25\% or 50\% of the full state of the system is observed in data assimilation framework.  

\begin{table}[htbp]
\caption{Quantitative assessment of different neural network parameterizations for subgrid scale processes using the total root mean square error given by Equation~(\ref{eq:rmse}).}
\begin{tabular}{p{0.4\textwidth}p{0.05\textwidth}}\\
\hline\noalign{\smallskip}
\textbf{Framework} & $\text{RMSE}$  \\\hline\noalign{\smallskip}
\underline{\emph{Only neural network parameterizations}} &  \\
ANN-3 & $3.38$ \\ 
ANN-5  & $3.73$ \\ 
ANN-7  & $3.77$ \\ 
CNN  & $3.79$ \\{\smallskip} 

\underline{\emph{Only data assimilation}} &  \\
No parameterizations ($m=9$) & $5.11$ \\ 
No parameterizations ($m=18$) & $4.30$ \\ 
No parameterizations ($m=36$) & $3.92$ \\{\smallskip} 

\underline{\emph{Neural network parameterizations with data assimilation}} &  \\
ANN-5 ($m=9$) & $0.52$ \\ 
ANN-5 ($m=18$) & $0.53$ \\ 
CNN ($m=9$) & $2.13$ \\ 
CNN ($m=18$) & $2.20$ \\ 
\hline
\end{tabular}
\label{tab:rmse}
\end{table}

\begin{figure*}[htbp]
\centering
\mbox{\subfigure{\includegraphics[width=0.95\textwidth]{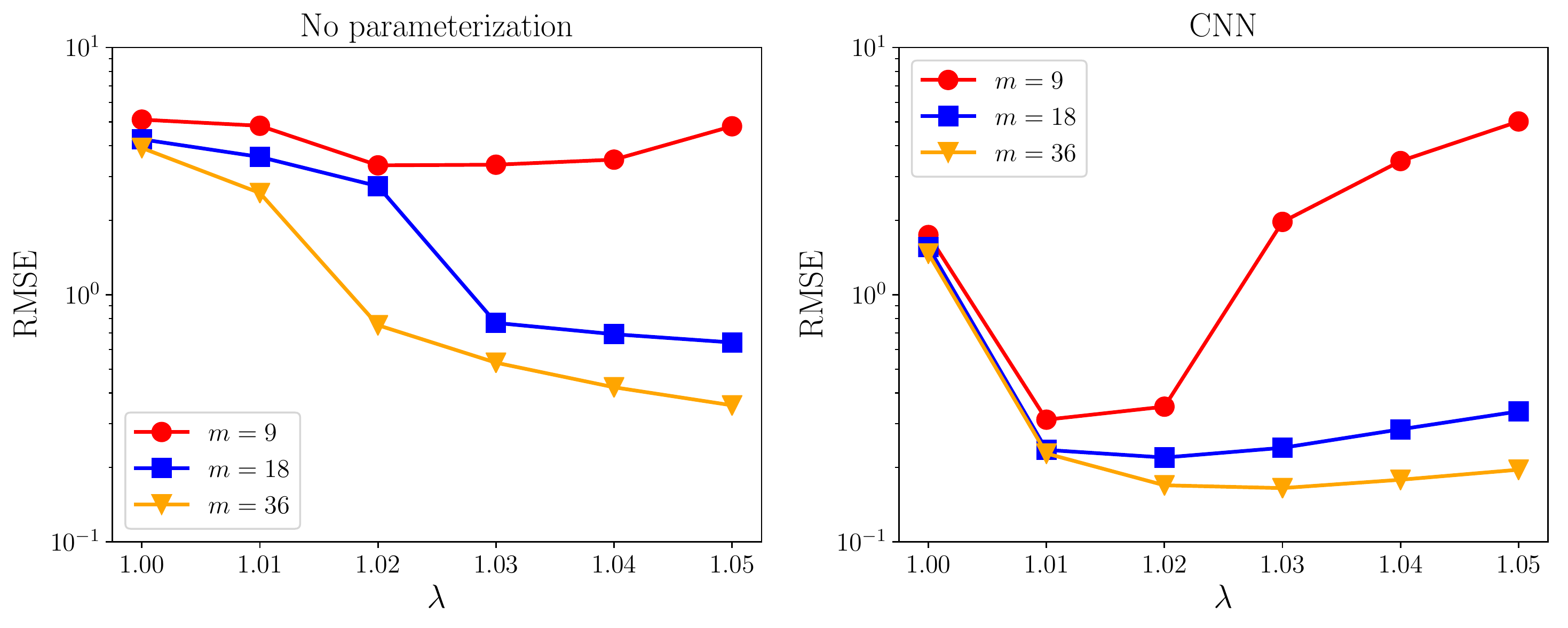}}}
\caption{The root mean squared error for different values of the inflation factor for three sets of observations. The number of ensembles is kept fixed at $N=30$ for all sets of observations.} 
\label{fig:ms_lambda}
\end{figure*}

\begin{figure*}[htbp]
\centering
\mbox{\subfigure{\includegraphics[width=0.8\textwidth]{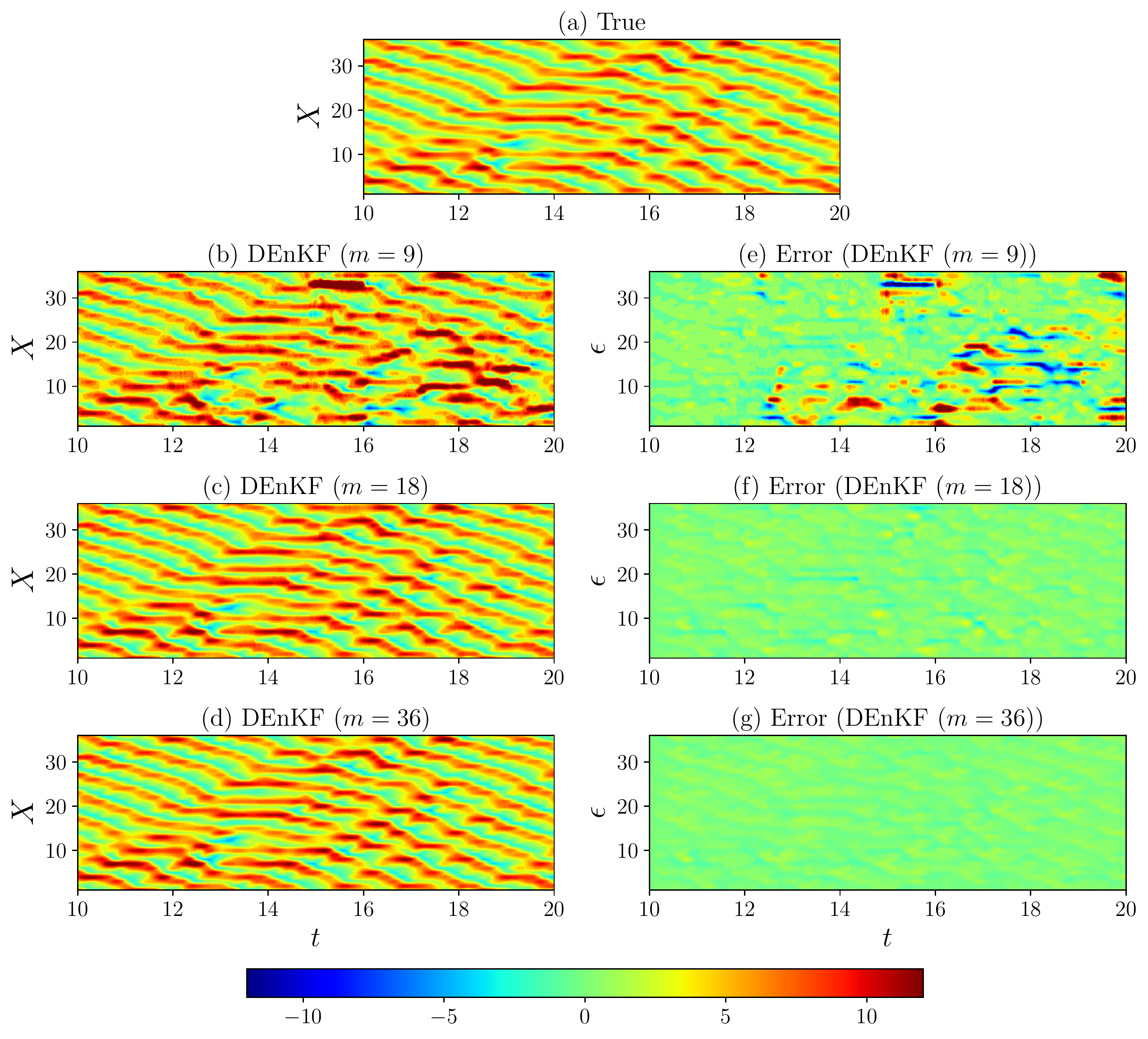}}}
\caption{Full state trajectory of the multiscale Lorenz 96 model with no closure for subgrid processes and for the inflation factor $\lambda=1.03$. The observation data for the DEnKF algorithm is obtained by adding measurement noise to the exact solution of the multiscale Lorenz 96 system.} 
\label{fig:l_nc_da_lambda}
\end{figure*}

\textcolor{rev1}{We highlight here that in the previous numerical experiment with the truncated model, we assumed that our forecast model is a true model. However, often the forecast models in DA are imperfect, and the model error introduced due to truncation of the sub-model is usually either modeled using the Gaussian noise or covariance inflation. Indeed, the ensemble Kalman filter framework is very well established and, to the extent that if modeling errors can be represented as zero-mean with a simple correlation structure, then the DA is very effective at correcting model errors. For example, \citet{brajard2020combining} utilized a Gaussian noise with zero mean and a certain value of standard deviation (optimized by tuning experiments) to account for the model error arising due to truncation of the parameterizations in a two-level Lorenz system. Similarly, \citet{attia2018optimal} proposed a variational framework for adaptive tuning of inflation and localization parameters and demonstrated its successful performance for a two-level Lorenz system. For a fair comparison with neural network-based parameterizations, we repeat the numerical experiments with the truncated model for different values of inflation factor. We keep the number of ensembles fixed at $N=30$, and the inflation factor is varied from 1.0 to 1.05 with an increment of 0.01. Figure~\ref{fig:ms_lambda} reports the RMSE for the truncated model for different inflation factors, and we can notice that with the proper choice of inflation factor and sufficient observations, the truncated model can also predict the true state of the two-level Lorenz system. In contrast to Figure~\ref{fig:l_nc_da}, Figure~\ref{fig:l_nc_da_lambda} depicts the full state trajectory of the two-level Lorenz system estimated using the DEnKF algorithm with the inflation factor $\lambda=1.03$ for three sets of observations. Overall, the results presented in Figure~\ref{fig:ms_lambda} suggest that the true state of the two-level Lorenz system can be determined when more than 50\% of the state is observable, and a proper value of inflation factor is employed for the DA with the truncated model. Moreover, the prediction of the true state of the system with neural network-based parameterization can be further improved by applying the inflation and the RMSE for different values of the inflation factor for CNN-based parameterization model are also shown in Figure~\ref{fig:ms_lambda}. It can be clearly seen that there is a significant accuracy gain by adding the CNN based parameterizations for almost all configurations. }

\textcolor{rev1}{
\subsection{Kraichnan turbulence}
We now characterize the performance of the EnKF algorithm, as described in Section~\ref{sec:da}, to estimate the state of the two-dimensional turbulence system when observation from high fidelity simulation are available. This test set-up is particularly challenging because of the modeling of unresolved scales in the LES solver, which is employed as the forecast model for two-dimensional turbulence. The performance of the EnKF algorithm is impacted by the choice of the model and the forecast model should be accurate enough for error control techniques like covariance inflation, covariance localization, stochastic forcing, etc. to work. As we will see, if the effect of unresolved scales are not modeled, even the EnKF algorithm with high value of the inflation factor does not improve the state estimate of the two-dimensional turbulence system.}

\textcolor{rev1}{
The governing equations for the two-dimensional turbulence are numerically solved using the second-order finite difference discretization. The nonlinear Jacobian term is discretized with the energy-conserving Arakawa \cite{arakawa1997computational} numerical scheme. A third-order total-variation-diminishing Runge-Kutta scheme is used for the temporal integration and a spectral Poisson solver is utilized to update streamfunction from the vorticity \cite{gottlieb1998total}. The computational domain is square in shape with dimensions $[0,2\pi] \times [0,2\pi]$ in $x$ and $y$ directions, respectively, and the periodic boundary condition is applied in both $x$ and $y$ directions. The training data for CNN is generated by carrying out the DNS at $\text{Re}=8000$ for two different initial conditions (independent of the truth model) on a grid resolution of $512 \times 512$ and then collecting total 800 snapshots (400 for each initial condition) between time $t=-2$ to $t=2$. The initial condition is assigned in such a way that the maximum value of the initial energy spectra occurs at wavenumber $K_p=10$. Further details of the randomization process for the initial condition can be found in related work \cite{san2012high}. The DNS data is coarsened to the $64 \times 64$ grid resolution using the spectral cutoff filter. The coarsened flow variables are then used to compute input features and labels for developing the data-driven subgrid-scale parameterization model as discussed in Section~\ref{sec:cnn}. Once the CNN is trained, it is deployed in the forecast model from time $t=0$ to $t=4$. We highlight that the data from time $t=2$ to $t=4$ is not seen during the training.  
}

\textcolor{rev1}{The predicted source term $\tilde{\Pi}$ has negative eddy viscosities embedded in it and needs to be post-processed before directly injecting it into the solver \cite{maulik2019subgrid}. The numerical stability is ensured during the \textit{a posteriori} deployment by truncating the learned source term $\tilde{\Pi}$ corresponding to negative numerical viscosities as follows  
\begin{equation}\label{eq:stability}
   \Pi = 
\begin{cases}
    \tilde{\Pi}, & \text{if } (\nabla^2 \bar{\omega})( \tilde{\Pi}) > 0 \\
    0,              & \text{otherwise}
\end{cases}
\end{equation}
In addition to truncating the source term corresponding to negative eddy viscosity, the truncation is also applied at points where the local eddy viscosity is greater than the local-average average eddy viscosity. This truncation scheme can be mathematically expressed as
\begin{equation}\label{eq:stability2}
   \Pi_{i,j} = 
\begin{cases}
    \tilde{\Pi}_{i,j}, & \text{if } \overbar{\nu}_{i,j} > \nu_{i,j} \\
    0,              & \text{otherwise}
\end{cases}
\end{equation}
where the eddy viscosity $\nu$ is computed as 
\begin{equation}
    \nu_{i,j} = \frac{\tilde{\Pi}_{i,j}}{\nabla^2 \bar{\omega}},
\end{equation}
and the local-averaged eddy viscosity $\overbar{\nu}_{i,j}$ is calculated using the mean filtering kernel of size $3 \times 3$. This additional truncation scheme given in Equation~\ref{eq:stability2} aids in preserving the statistical quantities like the kinetic energy spectra close to the DNS solution as compared to utilizing only negative eddy viscosity truncation scheme. In terms of the computational cost, the data-driven subgrid-scale parameterization model is significantly fast compared to the dynamic Smagorinksy model (DSM) \cite{pawar2019deep} and we observed up to 30\% reduction in computational speed in the \textit{a posteriori} runs with the CNN based closure model.
}

\textcolor{rev1}{
The `truth' solution for the data assimilation is obtained by solving the vorticity transport equation with a grid resolution $512 \times 512$ for the Reynolds number $\text{Re}=8000$ and then applying the spectral cutoff filter to get the filtered DNS solution on the coarse grid with resolution $64 \times 64$. Other methods like multigridding can also be adopted to relax the solution from fine grid to coarse grid \cite{popov2020stochastic}. The DNS solution is generated from time $t=-2$ to $t=0$ with $\Delta t = 1\times 10^{-3}$. The assimilation is started at time $t=0$ once the turbulence is developed and the initial transience from $t=-2$ to $t=0$ is discarded. The observations are assimilated at every $10^{\text{th}}$ time step of the forecast model. The synthetic observations are generated by sampling vorticity field at $32 \times 32$ (corresponding to 25\% of the full state of the system) equidistant points in $x$ and $y$ directions from the filtered DNS solution and then contaminating them with the Gaussian noise, i.e., $\mathbf{v}_k \sim {\cal{N}}(0,\mathbf{R}_k)$, where $\mathbf{R}_k = \sigma_b^2 \mathbf{I}$. We set observation noise at $\sigma_b^2=2$.  
}

\textcolor{rev1}{
The initialization of the ensemble members also plays an important role in the performance of the EnKF algorithm \cite{houtekamer2016review}, especially in the initial period of the DA. There are different ways that have been used for the initialization of ensemble members in DA of turbulent flows, such as, using the solution field separated by a certain time from the turbulent flow simulation \cite{colburn2011state}, adding random perturbation to mean flow solution \cite{da2018ensemble}. We initialize all ensemble ensemble members by adding a random perturbations drawn from ${\cal{N}} (0,\mathbf{P}_0)$ to the filtered DNS solution at time $t=0$. The initial covariance matrix is set at $\mathbf{P}_0=\sigma_0^2 \mathbf{I}$, where $\sigma_0^2=1$. 
}

\begin{figure*}[htbp]
\centering
\mbox{\subfigure{\includegraphics[width=0.95\textwidth]{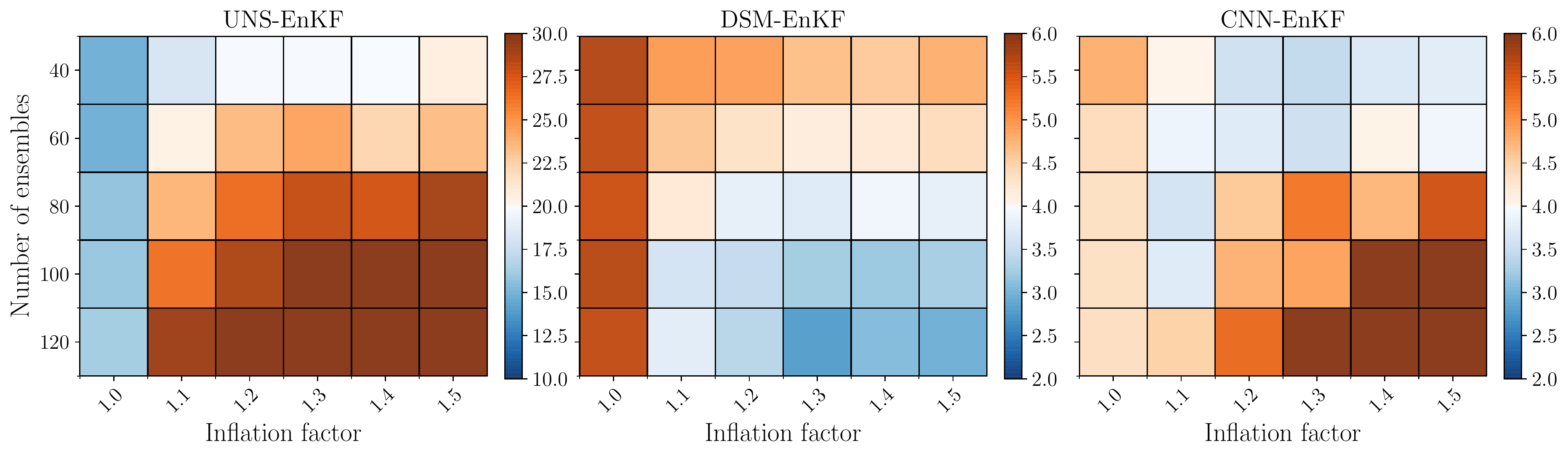}}}
\caption{Average RMSE of the compensated energy spectra for different combinations of the inflation factor and number of ensembles. The RMSE is averaged over the inertial range (i.e., from $K=8$ to $K=32$) considering data from $t=2$ to $t=4$.  } 
\label{fig:heatmap}
\end{figure*}

\textcolor{rev1}{We illustrate the performance of the EnKF algorithm for three different types of forecast models. The first forecast model is the unresolved numerical simulation (UNS), where subgrid-scale parameterization is completely discarded. In the second forecast model, the dynamic Smagorinsky model (DSM) \cite{germano1991dynamic,lilly1992proposed} is used for modeling the source term in LES simulation. The third forecast model consists of utilizing the CNN based subgrid-scale parameterization. For a fair comparison of the application of the EnKF algorithm to three different models, we run experiments with different combinations of the number of ensemble members and the inflation factor. The number of ensemble members is increased from 40 to 120 with an increment of 20 and the inflation factor is varied from 1.0 to 1.5 with an increment of 0.1. This gives us 30 different numerical experiments for each model. The performance of each numerical experiment is evaluated by computing the RMSE between the compensated kinetic energy spectra for the state estimated by the EnKF algorithm and the filtered DNS solution. The energy spectra is considered over the inertial range, i.e., between $K=8$ to $K=32$ and for 200 snapshots stored over the last half of the experiment timespan (from time $t=2$ to $t=4$). The RMSE results for these numerical experiments are shown in Figure~\ref{fig:heatmap}. Results in Figure~\ref{fig:heatmap} suggest that the RMSE for the UNS model is significantly higher than the DSM and CNN model. The solution field predicted by UNS has a significant error due to the truncation of subgrid-scale parameterization and an increase in the number of ensembles or the inflation factor does not seem to help improve the state of the system predicted by the UNS model. The average RMSE for both DSM and CNN models is of a similar magnitude. Moreover, Figure~\ref{fig:heatmap} indicates that the lower RMSE occurs at less number of ensembles for the CNN model with a moderate inflation factor (i.e., 1.2-1.3).   
}

\textcolor{rev1}{We evaluate the performance of different models through kinetic energy spectra calculation and second-order vorticity structure functions. We compute the  vorticity structure function using the formula given by \cite{grossmann1992structure} for two-dimensional turbulence and is shown below
\begin{equation}
    S_\omega (\mathbf{r}) = <|\bar{\omega}(\mathbf{x}+\mathbf{r}) - \bar{\omega}(\mathbf{x})|^2>,
\end{equation}
where $<>$ indicates ensemble averaging, $\mathbf{x}$ is the position on the grid, and $\mathbf{r}$ is certain distance from this location.
}

\begin{figure*}[htbp]
\centering
\mbox{\subfigure{\includegraphics[width=0.95\textwidth]{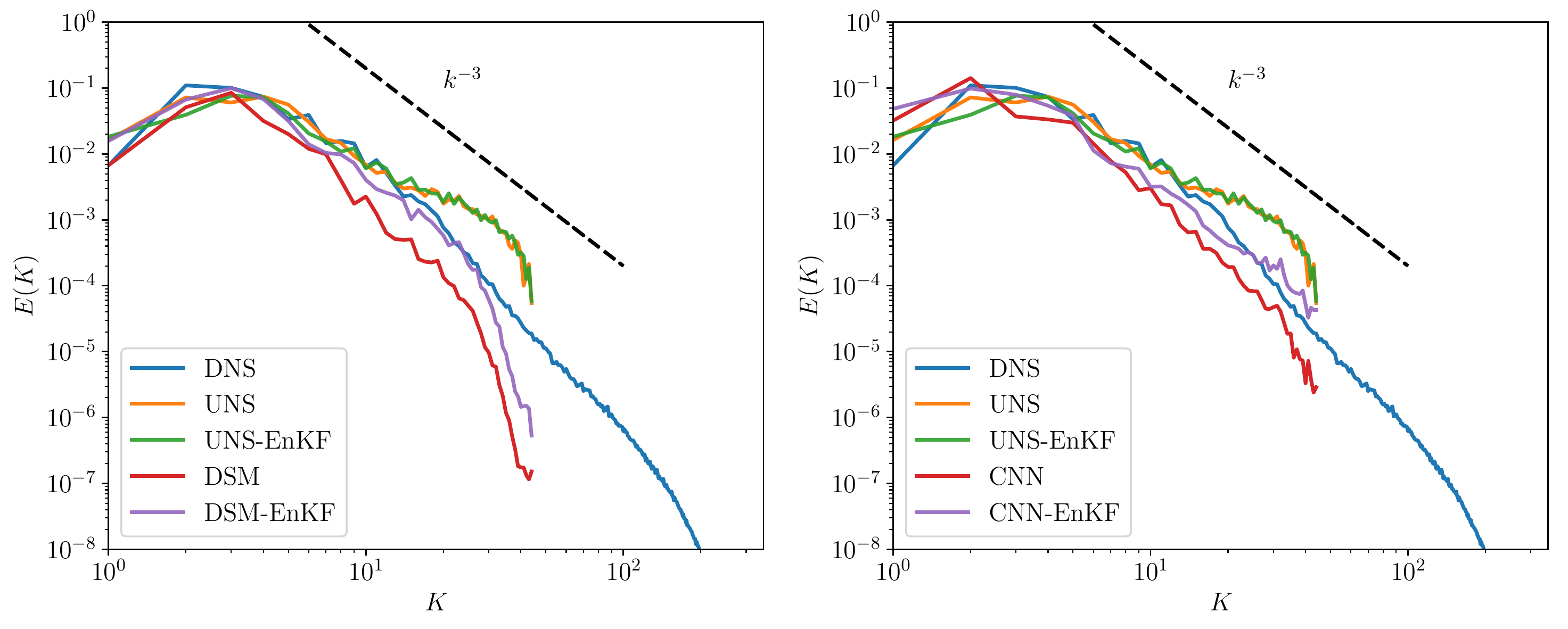}}}
\caption{\textit{A posteriori} kinetic energy spectra at $t=4$ at $N_x \times N_y = 64 \times 64$ grid resolution for different models. The number of ensembles and the inflation factor for the EnKF algorithm for different models corresponds to minimum value of the average RMSE between the compensated energy spectra for the filtered DNS solution and the solution predicted with different models. The EnKF related parameters are $N=40, \lambda=1.0$ for UNS-EnKF, $N=120, \lambda=1.3$ for DSM-EnKF, and $N=40, \lambda=1.3$ for CNN-EnKF.} 
\label{fig:ke_spectra}
\end{figure*}

\begin{figure*}[htbp]
\centering
\mbox{\subfigure{\includegraphics[width=0.95\textwidth]{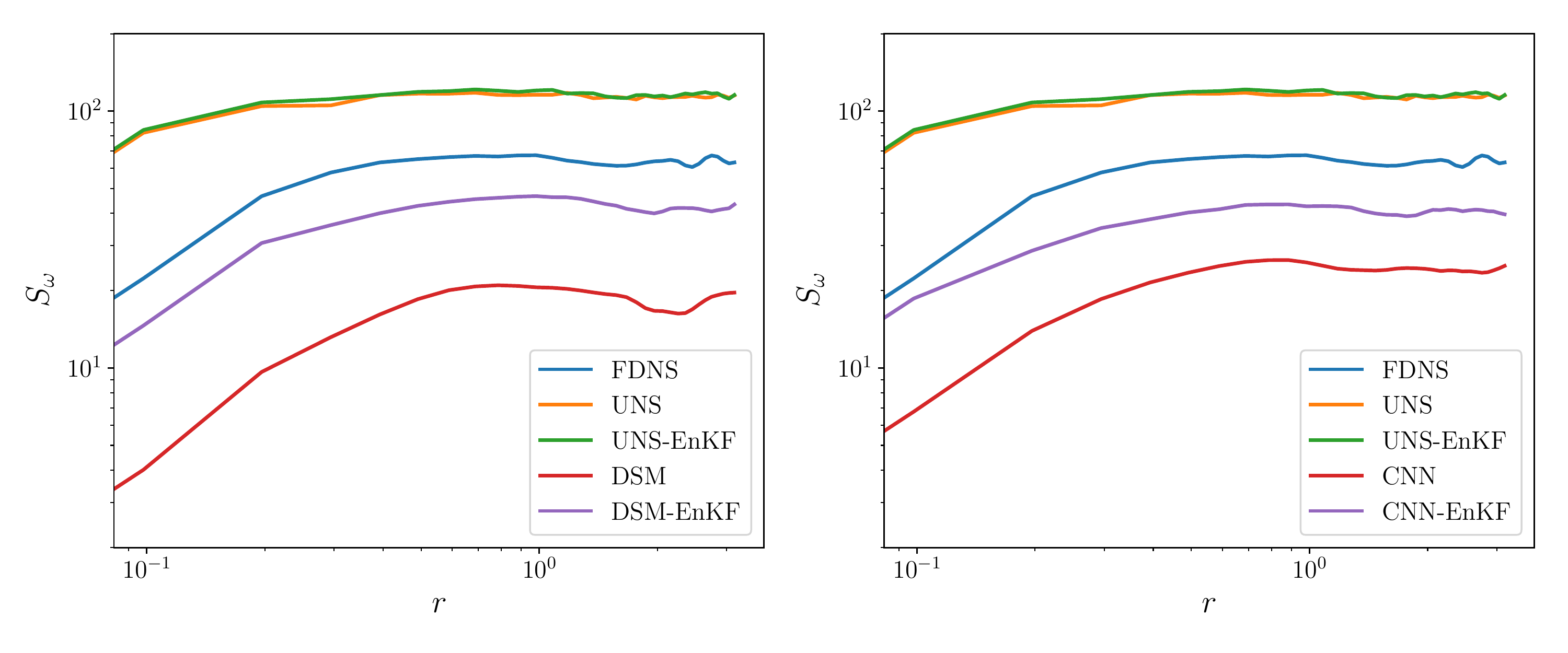}}}
\caption{\textit{A posteriori} second-order vorticity structure functions plotted against $r$ for $Re=8000$ at $N_x \times N_y = 64 \times 64$ grid resolution with different models used for the subgrid scale closure.} 
\label{fig:vort_struct}
\end{figure*}

\textcolor{rev1}{
Figure~\ref{fig:ke_spectra} displays the kinetic energy spectra at final time $t=4$ obtained with different models for $\text{Re}=8000$. We can observe that there is an accumulation of the energy near grid cutoff wavenumber in the case of the UNS model. The UNS-EnKF model is not able to correct the state estimate of the system due to very high noise in the forward model. We see an improvement in the energy spectra predicted by the DSM-EnKF and CNN-EnKF model compared to utilizing only the parameterization model. We note here that the kinetic energy spectra for the filtered DNS solution is identical to the DNS spectra till the grid cutoff wavenumber due to the use of a spectral cutoff filter. Figure~\ref{fig:vort_struct} depicts the second-order vorticity structure functions at final time $t=4$ where the evaluation with the FDNS shows that the EnKF is successful in improving the prediction of the vorticity structure function for both DSM and CNN model. We do not observe any improvement in the vorticity structure function prediction for the UNS model, which again emphasizes the importance of using an accurate forecast model in data assimilation.   
}

\section{Concluding Remarks}\label{sec:conclusion}

\textcolor{rev1}{The data-driven methods are successful in discovering model-free parameterizations from high-fidelity numerical simulations or experimental measurements and offers an alternative to parameterization models based on empirical or phenomenological arguments. The data-driven parameterization models are also computationally faster and are suitable for sequential data assimilation where multiple forward runs of a forecast model are required.} To this end, we introduce a framework to apply data assimilation methods to the physics-based model embedded with data-driven parameterizations to achieve accurate long-term forecast in multiscale systems.  We demonstrate that the forecasting capability of hybrid models can be significantly improved by exploiting online measurements from various types of sensor networks. Specifically, we use neural networks to learn the relation between resolved scales and the effect of unresolved scales (i.e., parameterizations). The deployment of the trained neural network in the forward simulation provides accurate prediction up to a short period and then there is a large discrepancy between true and predicted state of the system. To address this issue \textcolor{rev1}{ and to improve the long-term prediction}, we exploit the sparse observations data through data assimilation. 

We illustrate this framework for a two-scale variant of the Lorenz 96 model which consists of fast and slow variables whose dynamics are exactly known \textcolor{rev1}{and for Kraichnan turbulence where the parameterization model for unresolved scales is not known a priori}.  We obtain a considerable improvement in the prediction \textcolor{rev1}{for both test cases} by combining neural network parameterizations and data assimilation compared to employing only neural network parameterizations. We also found that including an ML based closure term seems to capture non-Gaussian statistics and significantly improve the forecast error. Based on our numerical experiments with data assimilation empowered neural network parameterizations, we can conclude that improving machine learning-based model prediction with data assimilation methods offers a promising research direction. \textcolor{rev1}{We also highlight that the inaccuracy associated with data-driven parameterizations can be tackled with data assimilation error control techniques like covariance inflation, covariance localization, stochastic forcing, etc.}     

Our future work aims at leveraging the underlying physical conservation laws into neural network training to produce physically consistent parameterizations. As the deep learning field is evolving rapidly, we can integrate modern neural network architectures and training methodology into our framework to attain higher accuracy. In the present framework, we employ the ensemble Kalman filter based algorithms for data assimilation. This algorithm gives accurate prediction when the uncertainty in model and observations follows a Gaussian distribution. We plan to investigate other data assimilation approaches like maximum likelihood ensemble filter methods that can handle the non-Gaussian nature of uncertainty in the mathematical model to get further improvement in the accuracy prediction. We will also test the present framework for more complex turbulent flows as a part of our future effort. Finally, we conclude by reemphasizing that the integration of data assimilation with hybrid physics-ML models can be effectively used for modeling of multiscale systems.    


\section*{Data availability}
The data that supports the findings of this study are available within the article. Implementation details and Python scripts can be accessed from the Github repository\cite{githubml}.


\begin{acknowledgements} This material is based upon work supported by the U.S. Department of Energy, Office of Science, Office of Advanced Scientific Computing Research under Award Number DE-SC0019290. O.S. gratefully acknowledges their support. 
Disclaimer: This report was prepared as an account of work sponsored by an agency of the United States Government. Neither the United States Government nor any agency thereof, nor any of their employees, makes any warranty, express or implied, or assumes any legal liability or responsibility for the accuracy, completeness, or usefulness of any information, apparatus, product, or process disclosed, or represents that its use would not infringe privately owned rights. Reference herein to any specific commercial product, process, or service by trade name, trademark, manufacturer, or otherwise does not necessarily constitute or imply its endorsement, recommendation, or favoring by the United States Government or any agency thereof. The views and opinions of authors expressed herein do not necessarily state or reflect those of the United States Government or any agency thereof.
\end{acknowledgements}

\appendix
\section{Validation of the Deterministic Ensemble-Kalman Filter} \label{app:validation}
In this Appendix, we provide results of data assimilation with the DEnKF algorithm for one level Lorenz 96 model. The one level Lorenz 96 model is given as 
\begin{align}
    \frac{d X_i}{dt} &= -X_{i-1} (X_{i-2} - X_{i+1}) - X_i + F, \label{eq:l96} 
\end{align}
for $i \in {1,2,\dots,36 }$ and $F=10$. The above model is completely deterministic as there is no parameterization of the unresolved scales. We use the similar settings as the two-level variant of the Lorenz 96 model for temporal integration using the fourth-order Runge-Kutta numerical scheme. The true initial condition is generated by integrating the solution starting from an equilibrium condition from $t=-5$ to $t=0$. For all ensemble members, we start with an initial condition obtained by perturbing the true initial condition with a noise drawn from the Gaussian distribution with zero mean and the variance of $1 \times 10^{-2}$. The observations are generated for data assimilation by adding a measurement noise from the  Gaussian distribution with zero mean and the variance of $\sigma_b^2 = 1$ (i.e, $\mathbf{R}_k = \mathbf{I}$) to the true state of the system. The observations are assumed to be available at every $10^{\text{th}}$ time step, similar to the two-level variant of the Lorenz 96 model. 

\begin{figure*}[htbp]
\centering
\mbox{\subfigure{\includegraphics[width=0.8\textwidth]{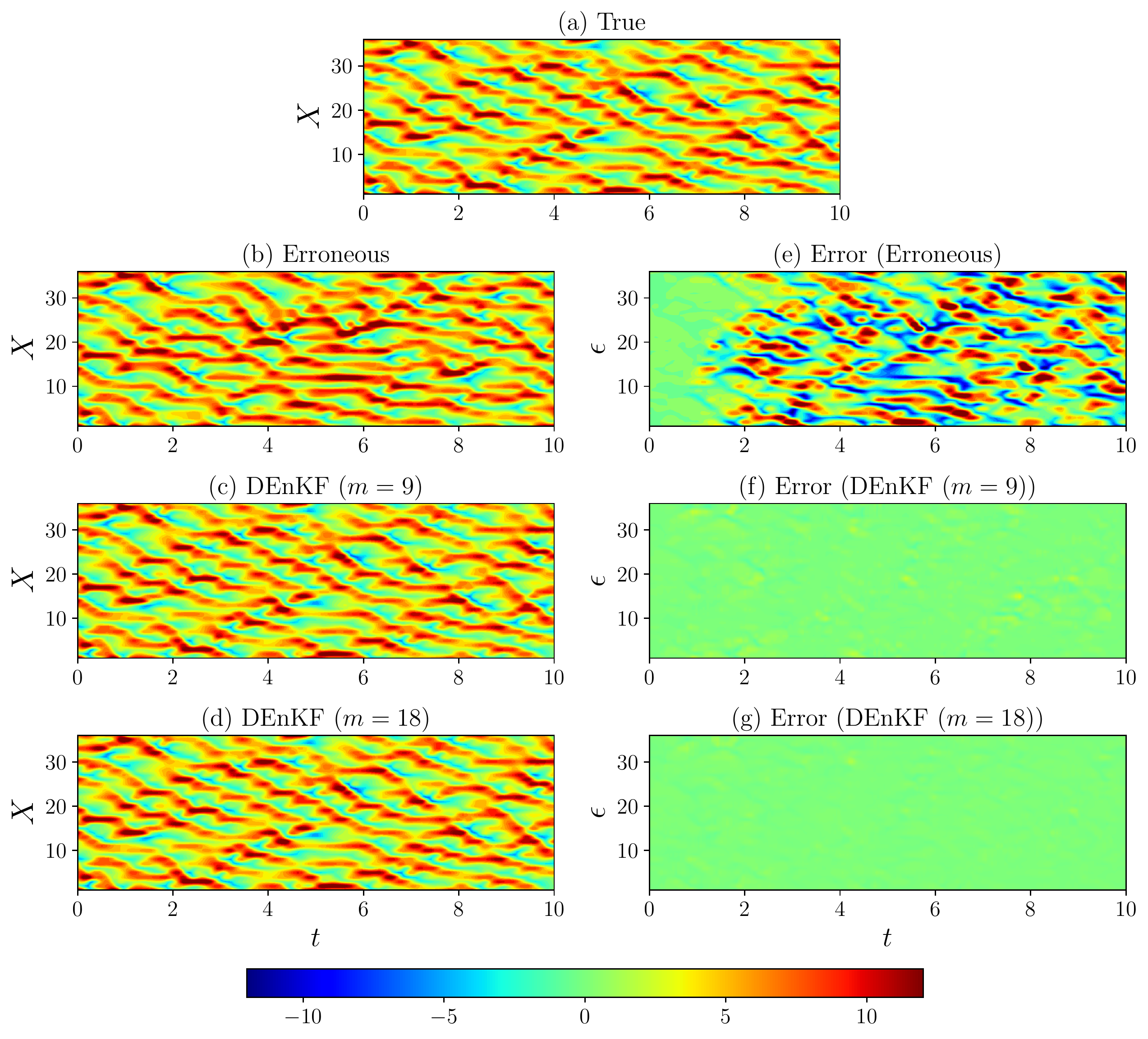}}}
\caption{Full state trajectory of the Lorenz 96 model with the the DEnKF algorithm.} 
\label{fig:l_da}
\end{figure*}

As depicted in Figure~\ref{fig:l_da}, we can conclude that the DEnKF can correct the erroneous trajectory even when only 9 observations are employed for data assimilation. As the amount of observations is increased to 18, we observe a reduction in the error. We reiterate here that, we have complete control over the model (since it is deterministic) in the numerical experiments with a one-level Lorenz 96 model. As we introduce fast scale variables, the evolution of slow variables in a two-level Lorenz 96 model is no longer deterministic and simple Kalman filter based algorithms might not be enough to give accurate prediction over a longer period.      

\bibliography{references}

\end{document}